\documentclass{scrartcl}

\usepackage[utf8]{inputenc}
\usepackage[iso]{datetime}
\usepackage{amssymb,amsmath,bm}
\usepackage{hyperref}
  \hypersetup{colorlinks,linkcolor=[rgb]{0.2,0.2,0.75},citecolor=[rgb]{0.3,0.5,0.5},urlcolor=[rgb]{0.4,0.4,0.6},breaklinks}
\usepackage[labelfont=bf]{caption}
\usepackage{graphicx}
\usepackage{siunitx}
\usepackage{subcaption}
\usepackage{authblk}
\usepackage{xcolor,soul}
\usepackage[backend=biber,style=ieee,citestyle=numeric,autocite=plain,sorting=none]{biblatex}

\newcommand{\coloredchange}[1]{\color{black}{#1}} 

\newcommand{\revised}[1]{{\color{purple}\coloredchange{#1}}}

\usepackage{xcolor}

\newcommand{\Romann}[1]{\uppercase\expandafter{\romannumeral #1\relax}}

\title{MHD simulations of formation, sustainment and loss of Quiescent H-mode in the all-tungsten ASDEX Upgrade}

\author[1,2]{Lorenz Meier}
\author[2]{Matthias Hoelzl}
\author[2]{Andres Cathey}
\author[1,3]{Guido Huijsmans}
\author[4]{Eleonora Viezzer}
\author[2]{Mike Dunne}
\author[1]{Jan van Dijk}
\author[4]{Diego José Cruz Zabala}
\author[2]{Karl Lackner}
\author[2]{Sibylle Günter}
\author[6]{ASDEX Upgrade Team}
\author[7]{EUROfusion MST1 Team}
\author[8]{JOREK Team}

\affil[1]{Eindhoven University of Technology, P.O. Box 513, 5600 MB Eindhoven, The Netherlands}
\affil[2]{Max Planck Institute for Plasma Physics, Boltzmannstr. 2, 85748 Garching b. M., Germany}
\affil[3]{CEA, IRFM, 13108 Saint-Paul-Lez-Durance, France}
\affil[4]{Dept. of Atomic, Molecular and Nuclear Physics, University of Seville, Avda. Reina Mercedes, 41012 Seville, Spain}
\affil[6]{See the author list of U. Stroth et al., Nucl. Fusion 62, 042006 (2022)}
\affil[7]{See the author list of B. Labit et al., Nucl. Fusion 59, 086020 (2019)}
\affil[8]{See the author list of M. Hoelzl et al., Nucl. Fusion 61, 065001 (2021)}

\date{\today}

\begin{document}

\maketitle

\begin{abstract}

Periodic edge localized modes (ELMs) are the non-linear consequences of pressure-gradient-driven ballooning modes and current-driven peeling modes becoming unstable in the pedestal region of high confinement fusion plasmas. In future tokamaks like ITER, large ELMs are foreseen to severely affect the lifetime of wall components as they transiently deposit large amounts of heat onto a narrow region at the divertor targets. Several strategies exist for avoidance, suppression, or mitigation of these instabilities, such as the naturally ELM-free quiescent H-mode (QH-mode). In the present article, an ASDEX Upgrade equilibrium that features a QH-mode is investigated through non-linear extended MHD simulations covering the dynamics over tens of milliseconds. The equilibrium is close to the ideal peeling limit and non-linearly develops saturated modes at the edge of the plasma. A dominant toroidal mode number of $n=1$ is found, for which the characteristic features of the edge harmonic oscillation are recovered.
The saturated modes contribute to heat and particle transport preventing pedestal build-up to the ELM triggering threshold. 
The non-linear dynamics of the mode, in particular its interaction with the evolution of the edge safety factor is studied, which suggest a possible new saturation mechanism for the QH-mode.
The simulations show good qualitative and quantitative agreement to experiments in AUG. In particular, the processes leading to the termination of QH-mode above a density threshold is studied, which results in the transition into an ELM regime. In the vicinity of this threshold, limit cycle oscillations are observed.

\end{abstract}
\section{Introduction}

The ITER tokamak~\cite{iterphysicsbasis} aims at operation in the high confinement mode (H-mode) to demonstrate high fusion rates and energy gains. This configuration is characterized by a steep pedestal pressure (large pressure gradient) close to the plasma boundary, which implies also a high bootstrap current density in this region. Such properties can render kink-peeling and ballooning modes linearly unstable, resulting in an edge localized mode (ELM) crash~\cite{elmreview} that expels large amounts of heat and particles from the plasma into the surrounding scrape-off layer, and ultimately to the divertor targets. After such an ELM crash, the plasma becomes MHD stable allowing the pressure pedestal to recover, resulting eventually in the next ELM crash. This periodically repeating process is foreseen to reduce the lifetime of divertor components in ITER as it gives rise to large transient heat loads that impinges onto a narrow area on the divertor targets~\cite{Loarte_2003,Eich_2017,Gunn_2017}. Multiple ELM cycles have previously been successfully simulated with the non-linear extended MHD code JOREK, emphasizing the role of plasma flows on the interplay of stabilising and destabilising eﬀects which eventually lead to an ELM crash~\cite{Cathey_2020}. If stabilizing flows are too weak to suppress MHD modes between ELM crashes, peeling-ballooning turbulence (``small ELMs'') arises instead of the isolated ELM crashes and prevents the pedestal build-up thereby preventing large type-I ELMs ~\cite{Oyama_2006,Labit_2019,XuGS_2019,Cathey2022,Viezzer_NME_2022} 

Close to the peeling stability boundary, in a regime with high bootstrap current (at low densities in present-day devices) another naturally ELM-free operational regime with good H-mode conﬁnement, denoted Quiescent H-mode (QH-mode), can be accessed~\cite{Burrell_2002,Suttrop_2003,Viezzer_2018}.
The QH-mode is characterized by the onset of an edge harmonic oscillation (EHO) which increases the heat and particle transport across the pedestal, such that the pedestal gradients remain below the ELM stability limit and the occurrence of ELMs is prevented. The EHO is typically observed to feature low toroidal mode numbers ($n=\{1\dots3\}$) and multiple higher harmonics~\cite{Burrell_2009}. It is conjectured to be a saturated kink/peeling mode, driven unstable by the high bootstrap current density, which allows the plasma to enter an ELM-free, quasi stationary state~\cite{Snyder_2007}; other works call it an infernal external mode~\cite{Medvedev_2006,Cooper_2016}. \revised{Historically, the QH-mode was first operated in the {DIII-D} tokamak at low pedestal density and high counter-current NBI injection~\cite{Burrell_2001}. After more than two decades of experiments it is now known that neither co-, nor counter-current NBI are necessary requirements for accessing QH-mode~\cite{Burrell_2009PRL}, and while the loss of QH-mode at high pedestal densities is a common feature observed in several machines, experiments in DIII-D have extended QH-mode operation to high density by careful tailoring of the plasma shape~\cite{Garofalo_2015_PoP,Solomon_2014_PRL}. What seems to be required for reliable QH-mode access is a sufficiently sheared ${\bm E\times \bm B}$ flow with a pedestal characterised as being near, but below, the peeling stability boundary (which is typically at low pedestal collisionality, i.e., low pedestal density in present-day machines) ~\cite{Burrell_2009,Burrell_2009PRL,Wilks_2018}. This picture has additionally been supported by theory and simulations~\cite{Snyder_2007,Chen_2016_Sim,XuGS_2017}.} This ELM-free regime is the subject of the present article, in which we present extended magneto-hydrodynamic (MHD) simulations of QH-mode access, sustainment and loss in ASDEX Upgrade (AUG)~\cite{Stroth2022} geometry using the JOREK code~\cite{JorekOverview}.

The article is structured as follows. In section~\ref{sec:Experiment}, the experimental background is introduced, followed by section~\ref{sec:setup} which presents the setup of our simulations. Section~\ref{sec:lr8} presents simulation results on the access and sustainment of QH-mode. Parametric dependencies regarding the effect of diamagnetic drift and the edge safety factor $q_{95}$ are presented in sections~\ref{sec:effectofdiamag} and~\ref{sec:VariationQ}, respectively. The termination of the QH-mode resulting from an increase in the plasma density is described in section~\ref{sec:QHtoELM}, followed by the description of an oscillation found close to the stability boundary~\ref{sec:LCO}. Conclusions and an outlook are finally given in section~\ref{sec:conclusions}.

\section{Experimental overview for ASDEX Upgrade discharge \#39279}
\label{sec:Experiment}

In this section, the experimental background of a relatively recent discharge in ASDEX Upgrade with all-tungsten walls that achieves transient QH-mode operation in spite of the metal walls is introduced. As this article cannot not provide a comprehensive review of experimental QH-mode observations in ASDEX Upgrade or other machines the reader is referred to Refs~\cite{Burrell_2002,Suttrop_2005,Viezzer_2018} and references therein for this purpose. 

Dedicated experiments at AUG recently established the QH-mode in a full metal environment for the first time~\cite{Viezzer_NME_2022,Viezzer_NF_2023}. Compared to the carbon wall QH-mode experiments, the establishment of the QH-mode in a tungsten wall environment is more challenging. This is due, in part, to the necessity of operating with a gas puff and core ECRH heating to prevent tungsten accumulation in the plasma core. Both ingredients are not necessarily beneficial to establish the QH-mode\revised{. The former because it leads to an increase of the pedestal density and the latter because it results in higher core $T_e$ than $T_i$, which makes accessing the QH-mode in AUG-W more challenging~\cite{Viezzer_NME_2022}}. At AUG, the QH-mode was obtained in two configurations, in a lower single null configuration with reversed field in which all NBI sources are counter-current, and in upper single null, forward field configuration, in which all NBI sources are co-current. Both feature the unfavorable ion $\nabla$B drift configuration. When operating in upper single null, density control is challenging as AUG does not feature a pump in the upper divertor. In this case, recycling plays \revised{an even} bigger role. One of the discharges from the upper single null experiments provides the basis for setting up the simulations carried out in this study and for various comparisons made between the simulations and experiments.

\begin{figure}[!ht]
	\begin{subfigure}[t]{0.68\linewidth}
	\centering
	\includegraphics[width=1.0\linewidth]{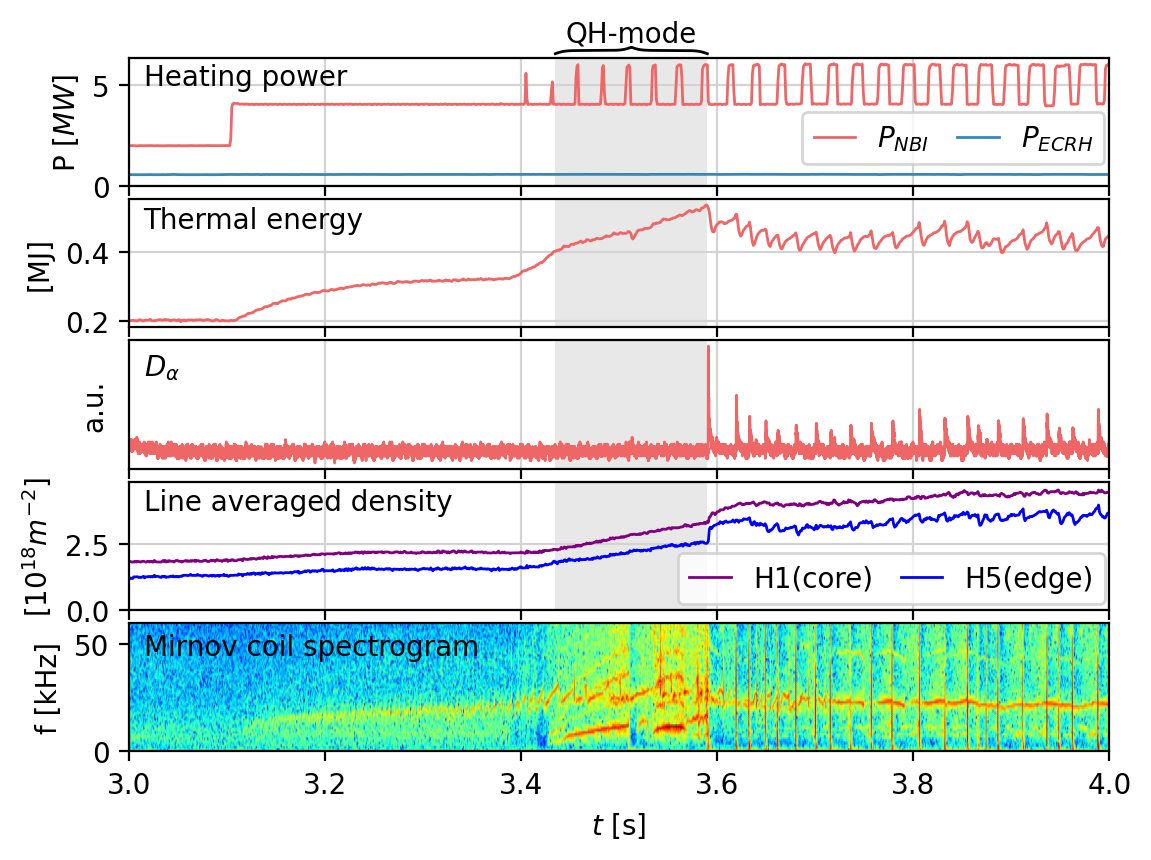}
	\caption[Overview of AUG discharge \#39279]{Overview of ASDEX Upgrade discharge \#39279. The QH-mode phase lasting from \SI{3.44}{\second} to \SI{3.59}{\second} is shaded in gray. \revised{Top to bottom:} NBI and ECRH heating power, total stored thermal energy, \revised{D-$\alpha$ signal,} line integrated density of two interferometer chords (core and edge), and a spectrogram of an outer midplane Mirnov coil, showing the emergence of the EHO with a dominant $n=1$ mode number and multiple higher harmonics. The signature visible throughout the entire time (beyond the QH-phase) corresponds to an unrelated core mode. In the ELMy H-mode following after the QH-mode phase, the periodic loss of stored thermal energy due to ELMs is visible. }
	\end{subfigure}
	\hfill
	\begin{subfigure}[t]{0.28\linewidth}
	\centering
	\includegraphics[width=\linewidth]{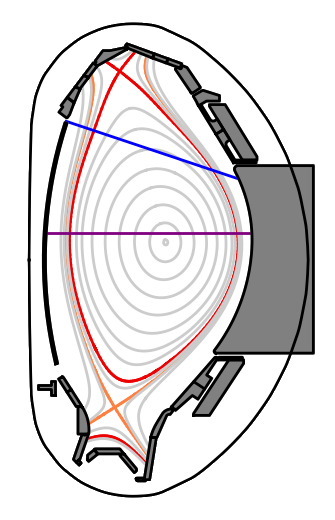}
	\caption[Equilibrium of AUG \#39279]{Reconstructed equilibrium of AUG \#39279 including PFCs and vacuum vessel. The inner separatrix is indicated in red, the outer in orange, while the interferometer chords H1 and H5 are indicated in purple and blue, respectively.}
	\end{subfigure}
	\caption[Overview of AUG discharge \#39279]{Overview of ASDEX Upgrade discharge \#39279}
	\label{fig:overview39279}
\end{figure}

As a basis for the simulations in this study, ASDEX Upgrade discharge \#39279, which features a transient QH-mode phase, was chosen. The discharge was carried out in forward field configuration and sustained QH-mode for about 150 milliseconds from \SI{3.44}{\second} to \SI{3.59}{\second}~\cite{Viezzer_NF_2023}. It features a toroidal magnetic field of $B_t= -2.5 \si{\tesla}$ and a plasma current of $I_P= 0.6 \si{\mega \ampere}$ (a positive sign here relates to the positive direction being counter-clockwise, when viewed from the top) with co-current neutral beam injection (NBI). The discharge was operated in an upper single null plasma configuration, close to a double null configuration with an elongation of $\kappa = 1.83$, an upper triangularity of $\delta_{\mathrm{upper}}= 0.44$ and a lower triangularity of $\delta_{\mathrm{lower}} = 0.31$. The magnetic geometry can be observed in Figure~\ref{fig:overview39279}(b).

The discharge was carried out in the unfavourable $\nabla B$ drift configuration (i.e., with the ion ${\bm B \times \nabla B}$ drift direction pointing away from the active X-point), which features a higher L-H power power threshold and allows us to achieve low density and high rotation already during the L-mode phase~\cite{ViezzerEPS}.

\begin{figure}[!ht]
	\centering
	\includegraphics[width=0.95\linewidth]{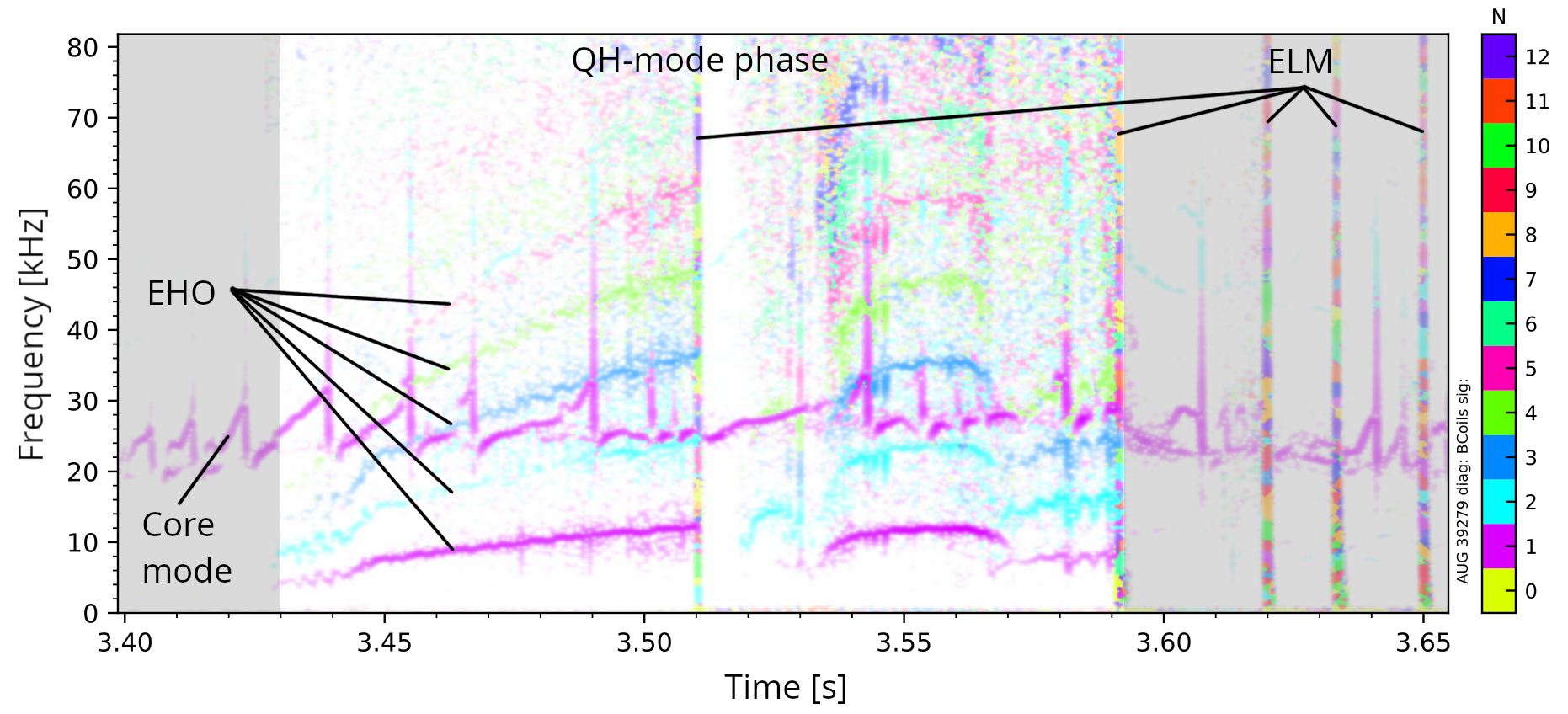}
	\caption[QH-mode phase cross-phaseogram]{Cross-phaseogram of several magnetic signals of the QH-mode phase from \SI{3.43}{\second} to \SI{3.59}{\second} and the subsequent ELMy H-mode, indicating the toroidal mode numbers. Multiple harmonics of the EHO can be seen to emerge in the QH-mode phase during which, despite the onset of the EHO, a small ELM occurs at \SI{3.51}{\second}. The QH-mode phase is terminated with a significantly larger ELM at \SI{3.59}{\second}. A $n=1$ core mode, unrelated to the QH-mode, can further be seen in the spectrum.}
	\label{fig:spectrogram39279}
\end{figure}

An overview plot for this discharge can be found in Figure~\ref{fig:overview39279}(a), which shows the heating power (combination of NBI \revised{plus NBI blips with increasing duration} and ECRH), the evolution of thermal energy, \revised{the D-$\alpha$ signal (which is an ELM monitor),} line-averaged density for two lines of sight, and a Mirnov coil spectrogram which allows to identify the Edge Harmonic Oscillation (EHO) \revised{together with a core mode that persists from $\sim3.15~\mathrm{s}$ to the end of the time window}. \revised{Both lines of sight together with the magnetic geometry are} shown in Figure~\ref{fig:overview39279}(b). A more detailed analysis of the mode spectrum can be found in Figure~\ref{fig:spectrogram39279}, which shows the emergence, toroidal mode number and frequency of the EHO during the QH-mode phase. 

In this discharge, during the QH-mode phase, the total thermal energy as well as the particle content increase, mainly through the rise of the density \revised{and ion temperature pedestals}, despite the emergence of the EHO. \revised{The increasingly long NBI blips cause the rise of the pedestals, which can be seen in fig.~\ref{fig:overview39279}(a).} The QH-mode phase \revised{is briefly interrupted for ${\sim20~\mathrm{ms}}$ at $3.51~\mathrm{s}$ coinciding with the end an NBI blip}. The EHO frequency shifts \revised{to larger frequencies with} time, likely due to the change in electric field well depth which deepens from $-65~\mathrm{kV/m}$ at \SI{3.475}{\second} to \SI{-110}{\kilo \volt \per \meter} at \SI{3.565}{\second}\revised{, which takes place due to the increase of the NBI power (as shown in the top of fig.~\ref{fig:overview39279}(a))}. The transient QH-mode phase eventually transitions to a type-I ELMy H-mode due to the increase of the pedestal density, which leads to the triggering of an ELM at \SI{3.59}{\second}.

\begin{figure}[!ht]
	\centering
	\includegraphics[width=0.7\linewidth]{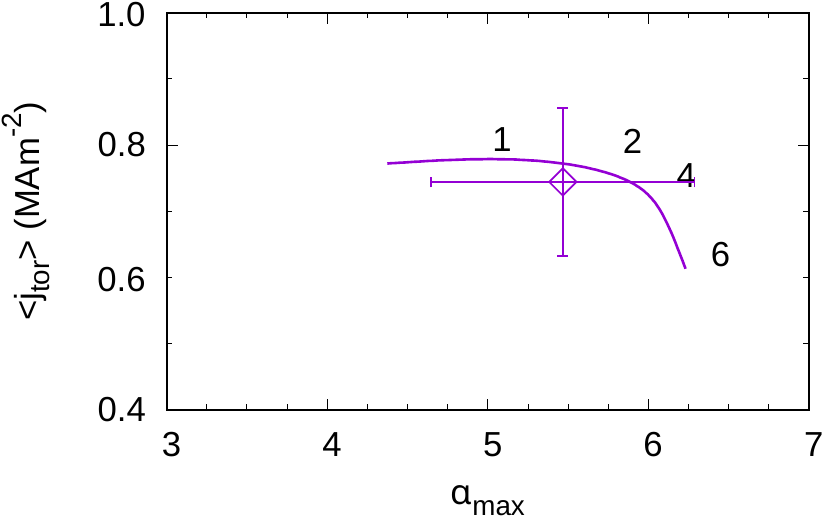}
	\caption[Linear stability analysis]{Linear ideal stability analysis of AUG \#39279 at \SI{3.475}{\second}. The stability boundary is indicated in a diagram of the normalized pressure gradient $\alpha$\protect\footnotemark and the flux surface averaged toroidal current density $\langle j_{tor}\rangle$. In the unstable region, the dominant toroidal mode number is indicated. It can be seen that the reconstructed equilibrium is close to the stability kink-peeling stability boundary as is commonly observed for QH-mode plasmas~\cite{Burrell_2009}.}
	\label{fig:Linear_MHD}
\end{figure}

Linear ideal stability analysis for the experimental equilibrium reconstruction at \SI{3.475}{\second} using the linear MHD code MISHKA-1 \cite{MISHKA-1} shows that the equilibrium is very close to the ideal kink-peeling stability boundary, with dominant toroidal mode numbers of $n=1, 2$, consistent with expectations for QH-mode, and matching the measurements from magnetic pick-up coils, i.e., Figure~\ref{fig:spectrogram39279}. The stability diagram and the operational point of the reconstructed equilibrium can be found in Figure~\ref{fig:Linear_MHD}.

\footnotetext{The normalized pressure gradient is defined as ${\alpha = - \frac{2 \mu_0 R_0}{B^2} q^2 \frac{\mathrm{d} p}{\mathrm{d} r}}$ where $\mu_0$ denotes the magnetic permeability and $R_0$ denotes the major radius.}

\section{Simulation Setup}
\label{sec:setup}

The simulations presented in this paper were carried out with the non-linear extended MHD code JOREK, which was developed in particular for the investigation of large scale plasma instabilities such as ELMs and disruptions as well as their control or mitigation. It uses a ﬁnite element method with a fully implicit time stepping scheme to solve the non-linear visco-resistive MHD equations in realistically shaped tokamak geometries for both the open and closed ﬂux surface regions. For this study, a reduced single temperature MHD model was used. The bootstrap current and diamagnetic flows are both incorporated self-consistently, while \revised{an ideally conducting wall was considered. Choosing such fixed boundary conditions neglects any electro-magnetic interactions that may take place between plasma and conducting structures, which has been studied with the JOREK-STARWALL coupling~\cite{Hoelzl_2012_starwall} for QH-mode simulations of DIII-D plasmas. In such studies it was found that considering resistive wall boundary conditions resulted in only moderate differences in the linear growth rates and non-linear saturated state~\cite{Liu_2015}. Nevertheless, future work will need to include such effects to quantitatively predict parameter thresholds for QH-mode access.}  A detailed description of JOREK and the reduced MHD model can be found in Ref.~\cite{JorekOverview}.

\begin{figure}
	\centering
	\begin{subfigure}[t]{0.53\linewidth}
		\centering
		\includegraphics[width=1.0\linewidth]{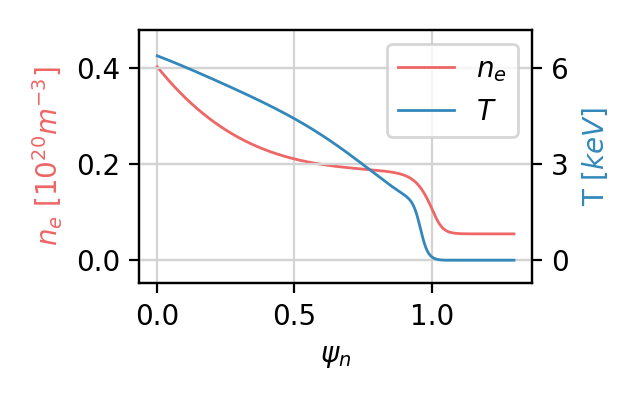}
		\caption[]{Initial electron density and temperature profiles used in the JOREK simulations.}
		\label{fig:init:rho}
	\end{subfigure}
	\hfill
	\begin{subfigure}[t]{0.46\linewidth}
		\centering
		\includegraphics[width=1.0\linewidth]{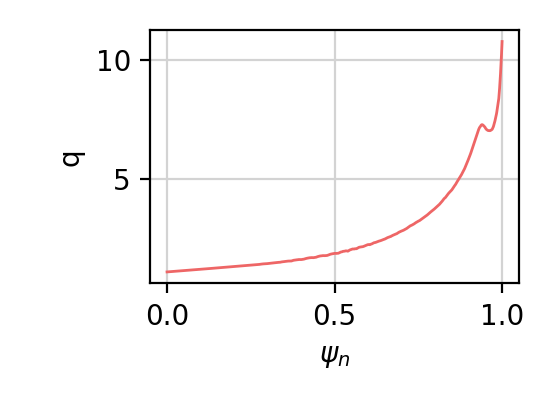}
		\caption[]{The initial q-profile of the equilibrium reconstruction with ${q_{95} = 7.1}$}
		\label{fig:init:q}
	\end{subfigure}
	\caption[Simulation input profiles]{Input profiles used for the simulation at nominal parameters.}
	\label{fig:init}
\end{figure}

As a starting point for the non-linear simulations, an equilibrium reconstruction from the \revised{early stage of the} QH-mode phase of AUG discharge \#39279 at $t = 3.475 \si{\second}$, obtained with the CLISTE code~\cite{CLISTE} is used. The input profiles for temperature and density, which were fitted to measurements taken with various diagnostics, as well as the $q$ profile are displayed in Figure~\ref{fig:init}.

\revised{The heat and particle sources applied in the simulation are chosen in combination with the \textit{ad-hoc} cross-field diffusion profiles such that the axisymmetric temperature remains roughly constant\footnote{in the experimental discharge $T_e$ remains constant while $T_i$ grows due to the NBI blips that are increasing in duration, but this effect is neglected in the simulations since we consider only a single-temperature treatment.} and that the density grows at a faster rate than the experimental evolution. Choosing the axisymmetric evolution of the density to exceed that of the experiment follows the approach at modelling pedestal instabilities described in Ref.~\cite{Cathey_2020}, which managed to simulate consecutive type-I ELMs.} Although the detailed deposition location of the heat source could be obtained by dedicated codes, \revised{they would require using sophisticated neoclassical and turbulent transport models, which goes far beyond the capabilities of the JOREK code at present. Instead,} the plasma is uniformly heated in the core of the plasma inside $\psi_n = 0.64$. The particle source \revised{is modelled as a Gaussian function at the edge, centered at $\psi_N = 1.0$ with a variance of $0.08^2$, such that it} deposits the majority of the particles in the edge region at $\psi_N > 0.9$ in order to maintain the pedestal shoulder. In the core, the particle source is kept constant and is roughly one order of magnitude smaller than the edge source. The input torque from the neutral beam injection is not considered for this simulation. \revised{The parallel velocity is instead set by the boundary conditions which enforce it to be equal to the ion sound speed at the divertor targets. The effect of the boundary condition onto the parallel velocity is then propagated to the entire SOL and inside of the separatrix. Investigating the role of $\boldsymbol{E} \times \boldsymbol{B}$ shear and input torque in a systematic way is left for future work since this is an extremely nuanced question. For the present modelling, the omission of input torque was deliberate in order to understand other control parameters (primarily pedestal density evolution and edge safety factor) independent of rotation and its shear as much as possible.}

The resistivity used in the simulations is specified by its core value and follows the Spitzer temperature dependency (${\eta \propto T^{-3/2}}$ ). The experimental resistivity can be determined by the Spitzer expression plus corrections from neoclassical effects and effective charge ($Z_{eff}$) \cite{Wesson}. For $\psi_N = 0.95$ this leads to an experimental resistivity of $1.1 \times 10^{-7}\, \si{\ohm \meter}$ while the resistivity in the simulation in that location is $9.9 \times 10^{-7}\,\si{\ohm \meter}$. The resistivity in the pedestal region is thus taken to be about a factor of $10$ larger than in the experiment. \revised{In contrast to JOREK simulations that investigate small-ELMs or the enhanced D-alpha H-mode in AUG which manage to simulate the non-linear MHD activity at realistic resistivity~\cite{Cathey2022,Cathey_2023_NF}, these QH-mode simulations probe experimental conditions at lower pedestal resistivity (roughly by a factor of 10 with respect to the aforementioned references) which make it much more complicated to consider realistic resistivity values. The exact impact of running with such larger resistivity onto the dynamics of the non-axisymmetric perturbations cannot be determined without running dedicated simulations. However, since the instabilities that are linearly unstable for the present simulations hold the mode structure of kink-peeling modes rather than ballooning modes, the effect of resistivity is not foreseen to be as important as it is for Refs.~\cite{Cathey2022,Cathey_2023_NF}.}

The bootstrap current is self-consistently evolved in time according to Ref.~\cite{Sauter} in all simulations presented here by a current source term that depends on the local plasma parameters. Diamagnetic drift terms and the neoclassical radial electric field are self-consistently included in all the simulations such that the ion fluid velocity is approximately at rest in the pedestal region~\cite{Orain2013,Viezzer_2014}. In a few selected simulations, these flow terms are artificially scaled to assess their impact onto the access of QH-mode. 

The flux aligned finite-element grid is constructed to have $161$ elements in poloidal direction, $105$ elements in radial direction inside the separatrix, $18$ radial elements outside the separatrix on the high field side and $20$ radial elements outside the separatrix on the low field side. The size of the elements is not uniform but is adjusted such that the pedestal region, where the MHD activity is situated, has the highest resolution with a radial finite-element width of $1$ to \SI{2}{\milli\meter}. For numerical stability reasons and to save computational costs, Taylor-Galerkin stabilisation is used, as well as hyper-diffusion, -viscosity and -resistivity. Via parameter scans, it is confirmed that these terms do not affect the physical results in the pedestal region. 

Based on the Grad-Shafranov equilibrium obtained directly in JOREK, all variables are initialized on the grid and the simulation is initially carried out axisymmetrically for \SI{0.2}{\milli \second} such that scrape-off layer flows can establish consistently with the Bohm boundary conditions. Thereafter, the simulation is continued non-axisymmetrically including six toroidal harmonics $n=1-6$ for most of the simulations. For the number of toroidal harmonics, the time step size and the poloidal resolution of the grid, parameter scans were performed to confirm that results are converged. 


\section{Simulations of QH-mode at experimental conditions}
\label{sec:lr8}

Using the non-linear MHD code JOREK, simulations of ASDEX Upgrade discharge \#39279, which features a \SI{150}{\milli \second} long QH-mode phase~\cite{Viezzer_NF_2023}, have been performed with parameters as outlined in section \ref{sec:setup}.  Since the model used does not include small-scale turbulence, JOREK cannot predict the background transport such that ad-hoc heat and particle diffusion coefficient profiles have to be provided. \revised{As mentioned in the previous section, these coefficients are chosen in combination with the source profiles in order to achieve an approximately constant temperature pedestal with a steepening density pedestal. The steepening of the density pedestal (in the absence of any perturbations) takes place at a slightly faster rate than the experimental evolution (in the presence of the EHO).}

\begin{figure}
	\centering
	\includegraphics[width=1.0\linewidth]{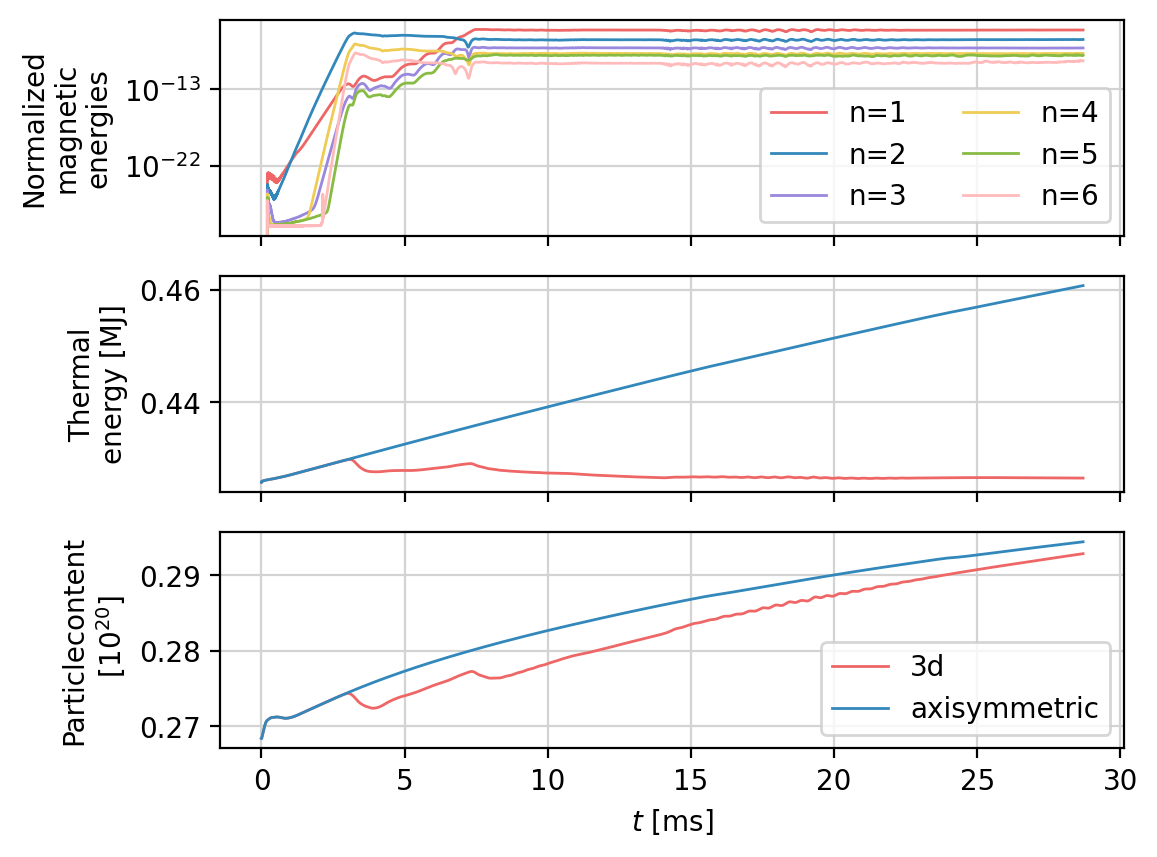}
	\caption[Evolution of magnetic energies, particle content and thermal energy during saturation]{Evolution of magnetic energies, particle content in the plasma and thermal energy in the plasma for a simulation including the toroidal modes ${n=\{0\dots6\}}$. The evolution in the absence of MHD modes is shown by plotting an axisymmetric case for comparison. The $n=1$ and $n=2$ modes are linearly unstable and grow initially. After about \SI{2}{\milli \second} the modes n = \{3\dots6\} are driven by n=1 and n=2 and start growing as well, eventually leading to a saturation of the n=2 mode at \SI{3}{\milli \second} and n=1 at \SI{7.5}{\milli \second}. Thereafter, the magnetic energies remain virtually stationary for at least \SI{20}{\milli \second}. The onset of the mode has an important influence on the heat transport and thereby on the thermal energy, however only a small influence on the particle transport.}
	\label{fig:macro_lr8}
\end{figure}

The magnetic energies, as well as the particle content and the thermal energy of this simulation are shown in Figure~\ref{fig:macro_lr8}\revised{; the latter also} in comparison to an axisymmetric simulation (in which no mode can emerge such that transport is purely driven by the background diffusion). Initially, $n=2$ and $n=1$ kink-peeling modes (KPMs) are linearly unstable \revised{in the JOREK simulations. These unstable modes sit within the error bars corresponding to the ideal linear picture of Figure~\ref{fig:Linear_MHD}}. The $n=2$ mode initially has a larger growth rate than the $n=1$ mode. Via three wave coupling~\cite{Krebs_Coupling}, the $n=2$ mode eventually starts to couple to the $n=4$ and $n=6$ mode harmonics and drives them from about \SI{2}{\milli \second} onwards. Shortly thereafter, the $n=2$ mode and associated higher harmonics saturate at about \SI{3}{\milli\second}, while the $n=1$ mode with a smaller growth rate continues to grow. The $n=3$ and $n=5$ harmonics are also excited non-linearly. Eventually, $n=1$ saturates as well at about \SI{7.5}{\milli \second} at a slightly higher magnetic energy compared to $n=2$ and remains dominant for the rest of the simulation. After the saturation of the KPM, the plasma enters a quasi stationary state, where the magnetic energies remain constant (except for small fluctuations) for the duration of the simulation, i.e., for at least \SI{20}{\milli \second}.

The onset of the saturated kink-peeling mode has an influence on both the particle content and the thermal energy content by causing additional transport across the pedestal. It is observed that the mode causes additional transport at both saturation events (at ${\sim3~\mathrm{ms}}$ for the $n=2$ mode, and at ${\sim7.5~\mathrm{ms}}$ for the $n=1$ mode). However, in the saturated state, only the heat transport is significantly changed, with respect to the axisymmetric evolution. The particle transport, measured by the particle flux at the computational boundary, is increased by approximately \SI{3.5}{\percent} after the saturation, while the energy transport, also measured by the energy flux at the computational boundary, is enhanced by approximately \SI{30}{\percent}. The additional heat transport by the mode is caused mostly by parallel transport along the stochastic field lines in the ergodic region that forms just inside the separatrix. After the saturation, both the pedestal top pressure, as well as the pedestal top temperature are about \SI{10}{\percent} lower compared to the axisymmetric case without MHD activity, the pedestal top density, however, is only slightly reduced.

Note, that while the heat source is well defined, the particle source is set up somewhat arbitrarily in this simulation and the same holds for the perpendicular particle diffusion coefficients. Consequently, the relative amplitudes of background transport and mode-induced transport might not be captured fully consistently in this simulation. A more accurate assessment of this is left for future work, where the scrape-off layer model can be significantly improved by accounting for kinetic neutrals such that the density source is taken into account in a more self-consistent manner. Input from transport code simulations could also be used to reduce uncertainties regarding particle sources and diffusion.

\begin{figure}
	\centering
	\includegraphics[width=0.95\linewidth]{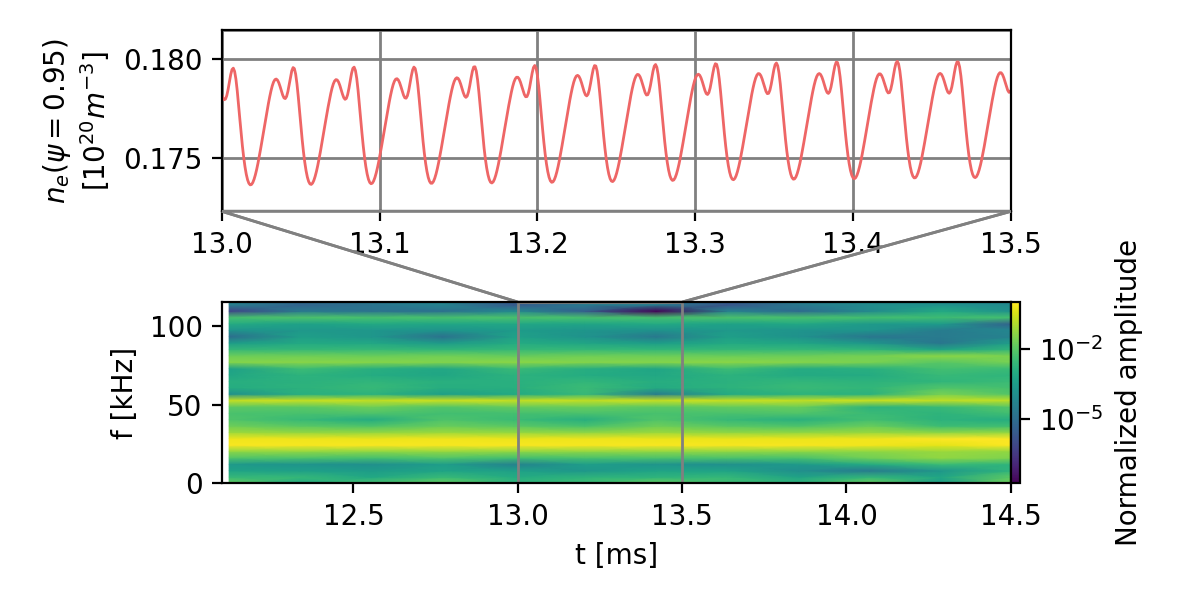}
	\caption[Oscillations at the outer midplane pedestal top]{Oscillations at the outer midplane pedestal top reminiscent of the EHO.
	The spectrogram shows the characteristic multipeaked spectrum with decreasing amplitude for the higher harmonics, indicating a coupling between the individual modes into a structure that is rigidly rotating.}
	\label{fig:EHO_spec}
\end{figure}

After the $n=1$ mode saturation, fluctuations in the pedestal can be observed in all simulated local quantities, in particular in the local temperature, density and pressure, which are caused by the toroidal rotation of the saturated KPM. The oscillations for the density at $\psi_\mathrm{n} = 0.95$ in the outer midplane and the corresponding spectrogram are displayed in Figure~\ref{fig:EHO_spec}. 

The oscillations show a clearly non-sinusoidal structure, qualitatively consistent with experimental observations of the EHO. We thus, call the oscillations originating from the saturated KPM an EHO. In the saturated state, the $n=1$ mode dominates the spectrum, while multiple higher harmonics with a constant frequency spacing and lower amplitude are visible. In this figure, only four harmonics are shown, however, all six toroidal harmonics included in this simulation follow the base frequency of the $n=1$ mode as ${f_n=n\,f_{n=1}}$. The conserved shape of the oscillations over time and the structure of multiple higher harmonics indicates a rigidly rotating structure formed by the coupled toroidal harmonics. The frequency of the EHO, as measured from the density oscillations, is found to be independent of the radial location in the pedestal.
The mode rotates in the electron diamagnetic direction, however, this cannot be directly compared to experimental observations due to the assumption of $T_e=T_i$ and the omission of the toroidal rotation source present in the experiment.

\begin{figure}
	\centering
	\includegraphics[width=0.95\linewidth]{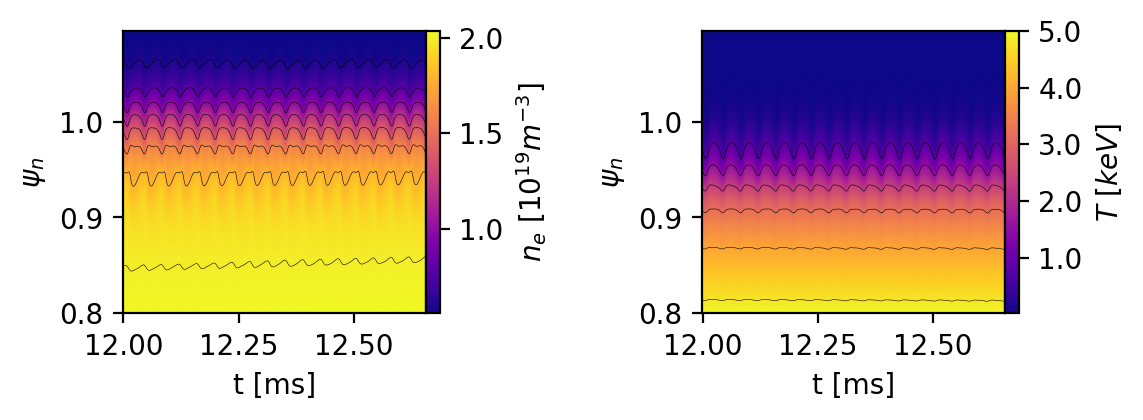}
	\caption[Oscillations in the pedestal]{Oscillations in the density and temperature pedestal reminiscent of the EHO are visualized by iso-lines in density and temperature. The non-sinusoidal structure of the oscillations is more pronounced in the density.
	}
	\label{fig:lr8_pedestaloscillations}
\end{figure}

The $n=1$ base frequency in the simulation is about \SI{26}{\kilo \hertz} \revised{for the entire simulation after the onset of the EHO}. Qualitatively, the observed fluctuations in the pedestal are in good agreement with the experimentally observed signals of an EHO and its frequency is close to typical frequencies observed experimentally~\cite{Paz_Soldan_2021}, however, the frequency is higher by about a factor of two compared to the EHO measured \revised{at the relevant point in time} in this particular discharge. Very likely, this is caused by the \revised{omission of the parallel rotation source in the simulations, while experimentally the toroidal velocity is measured to be rather large. The impact of including realistic sources of parallel rotation is left for future work that investigates closely the relationship between QH-mode and $\boldsymbol{E} \times \boldsymbol{B}$ shear and rotational shear. Qualitatively, however}, the present simulations show reasonable agreement with the experimental spectrogram shown in Figure~\ref{fig:spectrogram39279} \revised{even without the realistic parallel rotation, which would have a direct influence on the measured frequency of the saturated KPM.} Figure~\ref{fig:lr8_pedestaloscillations} illustrates the oscillations in the entire pedestal region by tracing iso lines of density and temperature over time. The non-sinusoidal structure of the oscillations can be seen to be more pronounced in the density than in the temperature. 

\begin{figure}[!ht]
	\begin{subfigure}{0.43\linewidth}
		\centering
		\includegraphics[width=1.0\linewidth]{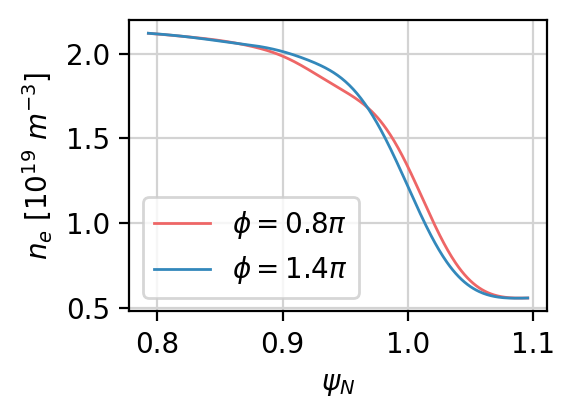}
		\caption[]{}
		\label{fig:lr8_torFTa}
	\end{subfigure}
	\begin{subfigure}{0.32\linewidth}
		\centering
		\includegraphics[width=1.0\linewidth]{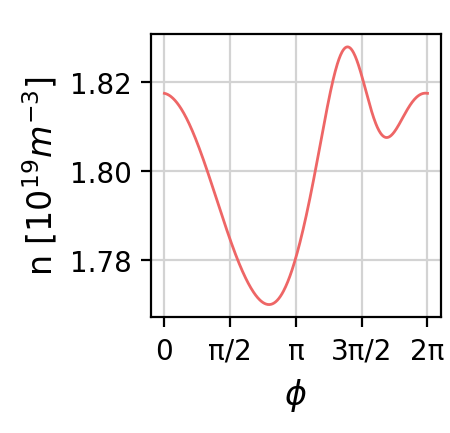}
		\caption[]{}
		\label{fig:lr8_torFTb}
	\end{subfigure}
	\begin{subfigure}{0.15\linewidth}
		\centering
		\includegraphics[width=1.0\linewidth]{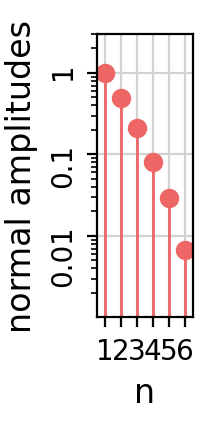}
		\caption[]{}
		\label{fig:lr8_torFTc}
	\end{subfigure}
	\caption[Variation of the density in the toroidal direction]{Variation of the density in the toroidal direction at \SI{20}{\milli\second}. (a) Profiles at toroidal angles $0.8\pi$ and $1.4\pi$ show a maximal displacement of the profile by approximately \SI{0.5}{\centi \meter} caused by the emerged mode. At the location of the mode on the pedestal top, the profile can be seen to be dented in. (b) Variation of the density along the toroidal angle in the outer midplane at $\psi_\mathrm{n} = 0.95$, indicating the toroidal localisation of the mode. (c) Fourier decomposition of the density fluctuation of (b), clearly dominated by n=1 with exponentially decaying amplitudes of the higher mode numbers.}
	\label{fig:lr8_torFT}
\end{figure}

\begin{figure}[!ht]
	\begin{subfigure}{0.25\linewidth}
		\centering
		\includegraphics[width=1.0\linewidth]{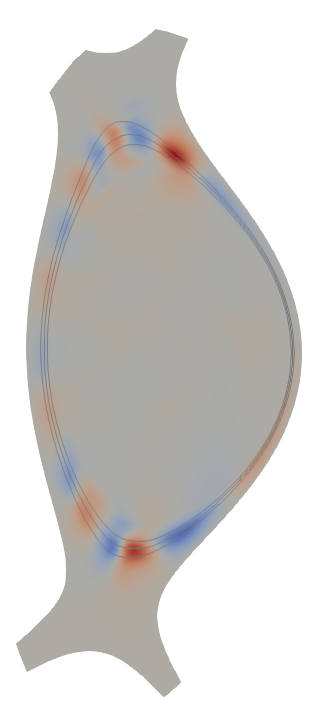}
		\caption[]{The perturbed poloidal magnetic flux, $\psi$}
	\end{subfigure}
	\hfill
	\begin{subfigure}{0.25\linewidth}
		\centering
		\includegraphics[width=1.0\linewidth]{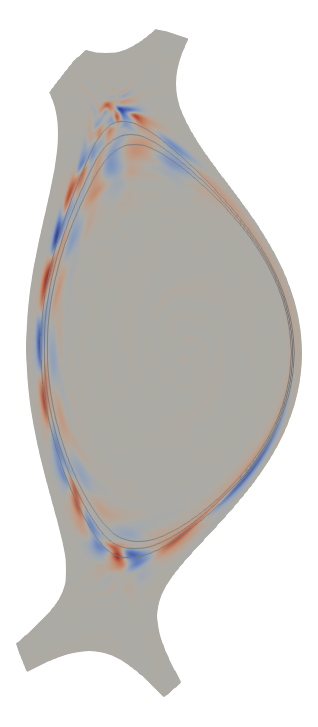}
		\caption[]{The perturbed density, $n_e$\\ }
	\end{subfigure}
	\hfill
	\begin{subfigure}{0.25\linewidth}
		\centering
		\includegraphics[width=1.0\linewidth]{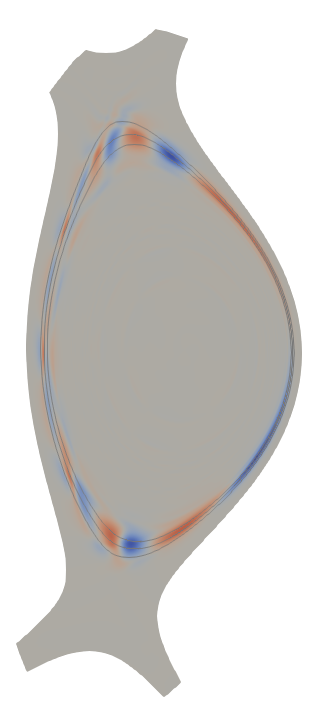}
		\caption[]{The perturbed temperature, $T$\\ }
	\end{subfigure}
	\caption[Poloidal mode structure]{Magnetic flux, density and temperature non-axisymmetric perturbations (including only $n\neq 0$ components) at \SI{12}{\milli\second}. 
	The KPM mode structure around $\psi \approx 0.97$ can be seen. The flux surfaces $\psi_\mathrm{n} = \{0.95, 0.97, 0.99\}$ are marked in black. Note that the density perturbations are larger on the HFS than on the LFS.}
	\label{fig:lr8_mode_structure}
\end{figure}

The toroidal structure of the mode can be investigated by plotting the fluctuations along the toroidal angle in the outer midplane, as shown in Figure~\ref{fig:lr8_torFTb}. It can be seen that the structure is dominated by the $n=1$ mode number, but also includes multiple higher harmonics as indicated by the Fourier decomposition given in Figure~\ref{fig:lr8_torFTc}. The phase relation between the toroidal harmonics is such that the resulting perturbation has an increased toroidal localisation. In Figure~\ref{fig:lr8_torFTa}, the density profiles are plotted at the toroidal angles $\phi=0.8\pi$ and $1.4\pi$, showing the maximal radial displacement of the pedestal over the course of one EHO period which is in this case approximately \SI{0.5}{\centi \meter}. In Figure~\ref{fig:lr8_mode_structure}, the $n \neq 0$ components of poloidal flux, density, temperature  are shown in a poloidal plane. This mode structure can be identified as a KPM with a maximal amplitude at the flux surface at around $\psi_\mathrm{n} = 0.97$.

\begin{figure}[!ht]
	\centering
	\includegraphics[width=0.95\linewidth]{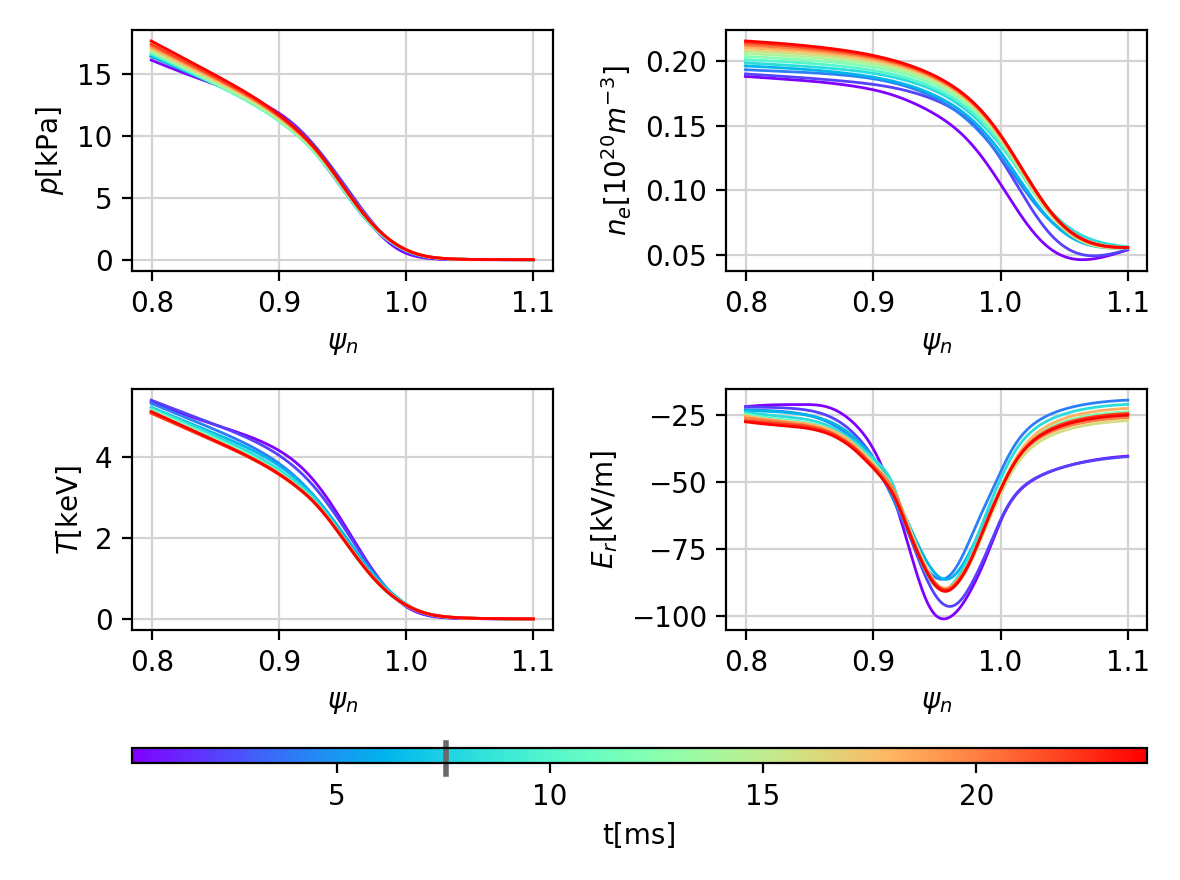}
	\caption[Evolution of the profiles during QH-mode]{Evolution of the toroidally averaged pressure, density, temperature and radial electric field profiles. Before the saturation of the $n=1$ mode, the profiles still evolve strongly. After the saturation (indicated in gray on the time scale), only a moderate evolution of the profiles is seen except for the density\revised{, which continues to steepen}. The $E_r$ well observed in the simulation is particularly deep as it is also seen in the experiment.}
	\label{fig:lr8_profileevolution}
\end{figure}

The evolution of the toroidally averaged (i.e., n=0 components) pedestal pressure, density, temperature and radial electric field is displayed in Figure~\ref{fig:lr8_profileevolution}. During the linear growth and saturation phase from \SI{0}{\milli \second} to \SI{7}{\milli \second}, the profiles evolve strongly, in particular $E_r$ adapts to the new saturated state, while the temperature pedestal slightly drops and the density increases keeping the pressure profile almost stationary. After the saturation, the density in the pedestal increases, like in the experiment, while the pedestal temperature slowly decreases mainly due to two effects: 1) parallel conduction due to a narrow ergodic region and 2) reduction of the heating power per particle. Over the entire evolution, the pressure profile and its gradient remain \revised{roughly} constant \revised{because the density increases while the temperature is reduced}. The radial electric field well is particularly deep in this case due to the low density and high ion temperature in the pedestal and is in good agreement with the experimentally measured depth of the $E_r$ well of ${-65 \pm 10 \si{\kilo\volt\per\meter}}$ \cite{Viezzer_NF_2023}, even though it is slightly radially displaced, having its minimum at ${\psi_\mathrm{n} = 0.99}$ experimentally while in the simulations it is at ${\psi_\mathrm{n} = 0.96}$. This discrepancy could potentially be a result of using the single temperature model, as it was seen in experimental measurements that the maximal ion temperature gradient (which influences the ion pressure pedestal and thereby the electric field well) is located at ${\psi_\mathrm{n} = 0.985}$ whereas the maximal electron temperature gradient is located further inwards at ${\psi_\mathrm{n} = 0.965}$. \revised{Unlike in the experiment, no deepening of the Er well can be observed in the simulations during the QH mode because the increasing toroidal rotation and $T_i$ pedestal top resulting from the NBI blips becoming longer are neglected.} 

\begin{figure}[!ht]
	\centering
	\includegraphics[width=0.95\linewidth]{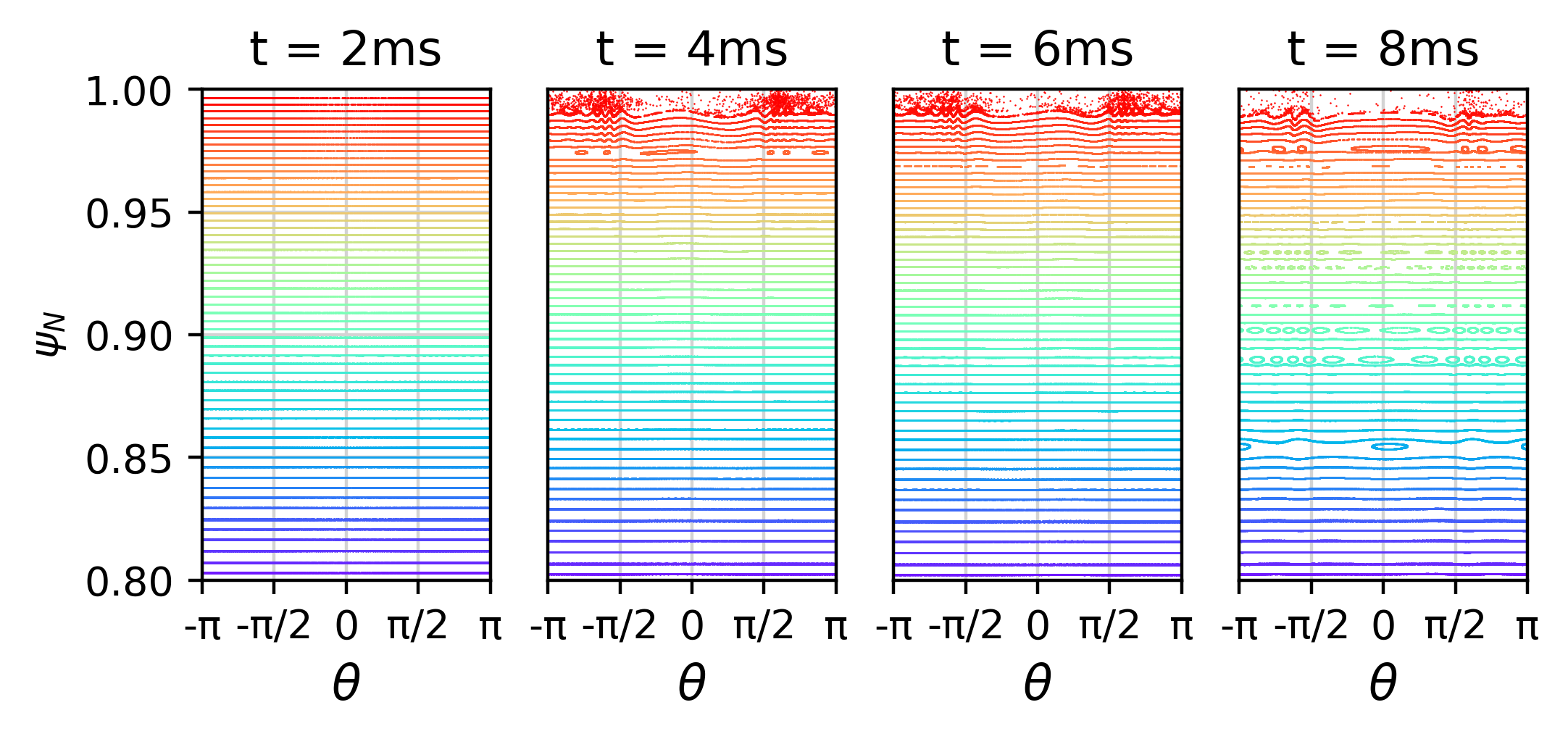}
	\caption[Poincaré plots of the saturation phase]{Poincaré plots of the early phase of the simulation. The plots correspond, from left to right, to the points in time before and after the saturation of the $n = 2$ mode and before and after the saturation of the $n = 1$ mode.}
	\label{fig:lr8_Poincare}
\end{figure}

Figure~\ref{fig:lr8_Poincare} shows Poincaré plots of the magnetic field structure for four points in time ($2, 4, 6$ and $8~\si{\milli\second}$) over the course of the growth and saturation of the mode. It can be seen that in the stationary state after the saturation of $n=1$ (at ${t=8~\si{\milli\second}}$), the magnetic surfaces remain closed up to a flux surface of approximately ${\psi_\mathrm{n} = 0.987}$. Outside this radius, an ergodic layer forms, leading to energy losses via parallel transport along the stochastic field lines. 

After the saturation, the mode structure remains virtually stationary for the entire simulated timescale of up to \SI{30}{\milli \second} with only minor oscillations in the mode amplitude. After $35~\mathrm{ms}$, the simulation was halted because of the computational expense, and because a \revised{linearly unstable} ${(m,n)=(2,1)}$ tearing mode \revised{which very slowly grows throughout the simulation became large enough} in the core that makes it harder to separate in virtual diagnostics between QH-mode effects and core mode influence (note that also the experiment shows sign of a core mode, as seen in Figure~\ref{fig:spectrogram39279}).

To test the relevance of higher mode numbers, for the exact same initial conditions, a simulation was carried out, which did not only include $n = \{0..6\}$ modes but $n = \{0..12\}$, such that also high $n$ ballooning modes could emerge if they were destabilised in this simulation. The linear mode spectrum did not change, however, as only $n=1$ and $n=2$ are linearly unstable. Barely any difference between the two simulations could be observed non-linearly, the dominant $n=1$ magnetic energy was reduced by only \SI{0.15}{\percent} while the $n=6$ magnetic energy was reduced by \SI{5.2}{\percent}. This shows that including $n = \{0..6\}$ modes, converged results are obtained justifying this choice for our QH-mode studies. For this reason, all further simulations presented in this article are done with the toroidal modes $n = \{0..6\}$ included.

An additional test was carried out, in which the resistivity was reduced by a factor of two. While details of the initial non-linear saturation and the EHO-induced transport differ slightly, the plasma forms a very similar non-linear state. Quasi-stationary modes with a dominant $n=1$ component establish just like in our original setup. Due to the large number of cases studied in this article, we use the original setup with the resistivity increased by a factor $10$ for the rest of our studies, which is computationally easier to handle.

\section{Effect of diamagnetic drift}
\label{sec:effectofdiamag}
\begin{figure}
	\centering
	\includegraphics[width=0.95\linewidth]{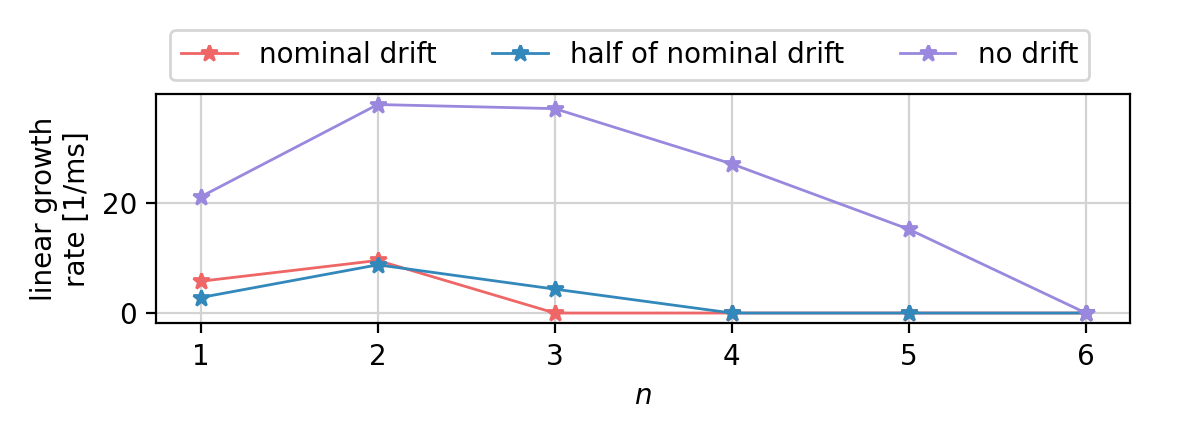}
	\caption[Linear growth rates dependant on diamagnetic drift]{Linear growth rates of the $n=\{1..6\}$ modes for simulations without diamagnetic drift (purple), with half of the nominal diamagnetic drift (blue) and with nominal diamagnetic drift (red).}
	\label{fig:diaggrowth}
\end{figure}

Diamagnetic drift effects play a significant role for the formation of QH-mode in the considered ASDEX Upgrade configuration as is shown and discussed in the following. This had been neglected or was not included fully self-consistently in previously published simulations of the QH-mode \cite{Liu_2017, Liu_2015, Pankin_2020}. Note, however, that in large machines like ITER, the diamagnetic drift effect will be much smaller than in the medium size tokamak ASDEX Upgrade. To better understand the influence of diamagnetic drift (and the related $\bm E \times \bm B$ flows), an additional set of simulations was carried out, in which the diamagnetic drift was switched off entirely or was reduced to half of its nominal value. \revised{For the JOREK simulations presented here, running without diamagnetic drift means that the radial electric field does not display the well in the pedestal which is characteristic of high-confinement regimes, as described in Ref.~\cite{Cathey_2021_PPCF}. Namely, the inclusion of the diamagnetic extension in JOREK produces the pedestal well in the $E_r$ profile. At half of the nominal drift, the depth of the $E_r$ is reduced by half. The cases with nominal, half, and no diamagnetic drift then have different ${\bm E\times \bm B}$ flow and shear, which has a direct effect on stability.} The linear growth rates for these simulations can be found in Figure~\ref{fig:diaggrowth}. 

If the diamagnetic terms are switched off, the growth rate of the still linearly dominant $n=2$ mode is about four times larger compared to the case with nominal diamagnetic drift terms. Non-linearly, the $n=2$ mode in the absence of diamagnetic drift results in a burst-like event and reaches a magnetic energy $6.5$ times larger than in the nominal case. Consequently, it causes significantly more transport across the pedestal. Shortly thereafter, the two cases are no longer comparable in a meaningful way as the case without diamagnetic drift develops a disruptive scenario with large islands inside the core and rapidly loses large parts of its energy content. This shows that the inclusion, respectively the non-physical exclusion, of the diamagnetic drift can change the linear and non-linear situation dramatically.

When the diamagnetic drift is half of the nominal value, the growth rate of the fastest growing $n=2$ mode is almost identical to the nominal case, but its non-linear evolution is more violent. Namely, the KPM can grow to a perturbation energy more than three times larger compared to the case with nominal diamagnetic drift and causes more transport across the pedestal. Note that the n=1 growth rate is higher in the nominal case compared to the simulation with reduced diamagnetic drift term. Thus, the plasma flows start to act destabilizing for the n=1 mode beyond a specific threshold. Modelling work with M3D-C1 and BOUT++ have found rotation and/or rotational shear to play a destabilising role for low-n modes as well~\cite{Chen_2016_Sim,XuGS_2017}. \revised{A similar observation has been reported with the MINERVA-DI code, which solves the linear ideal MHD stability of toroidally rotating plasmas and can account for the diamagnetic drift, but regarding destabilisation of higher-n ballooning modes upon including the realistic toroidal rotation profiles~\cite{Aiba_2016,Aiba_2018}. Such observation of realistic toroidal rotation profiles leading to destabilisation of high-n ballooning modes has not been obtained with JOREK.}

Cases with reduced diamagnetic drift are more closely comparable to QH-mode simulations in JOREK that were previously done without inclusion of the diamagnetic drift \cite{Liu_2015, Liu_2017}, which initially showed a burst like behaviour when the mode saturated, similarly to the one observed here. These publications had shown that a modification of the $\bm E \times \bm B$ velocity via an artificial source term strongly affects the growth and saturation of the KPM.

\section{Variation of the q-profile}
\label{sec:VariationQ}

To assess the influence of the edge safety factor ($q_{95}$) for the access to QH-mode, the q-profile is varied by scaling the toroidal magnetic field, which rigidly scales the full q-profile up and down without affecting the global magnetic shear. Similar variations have also been performed in experiments in the past in the form of a $B_t$ ramp during a discharge~\cite{Suttrop_2004}. In the simulation, individual cases are carried out with different values of the initial edge safety factor from $q_{95} = 6.4$ (where the q-profile in the core is just above $1$) up to $q_{95} = 8.15$, thus scanning a window of $-10\%$ to $+15\%$ around the nominal value of ${q_{95} = 7.1}$. Within the range covered by this scan, three discrete windows are found where QH-mode can be accessed. 
Outside these windows, the plasma displays either strong ELM-like bursts, or remains stable to all toroidal harmonics for long periods of time, such that the occurrence of large ELMs would be expected eventually. 
The following describes first how the linear growth rates change with the edge safety factor (Section~\ref{sec:VariationQLin}) and, thereafter, discusses the non-linear evolution of such cases (Section~\ref{sec:VariationQNonLin}).

\subsection{Linear growth-rate as a function of edge safety factor} \label{sec:VariationQLin}

\begin{figure}[h!]
	\centering
	\includegraphics[width=0.95\linewidth]{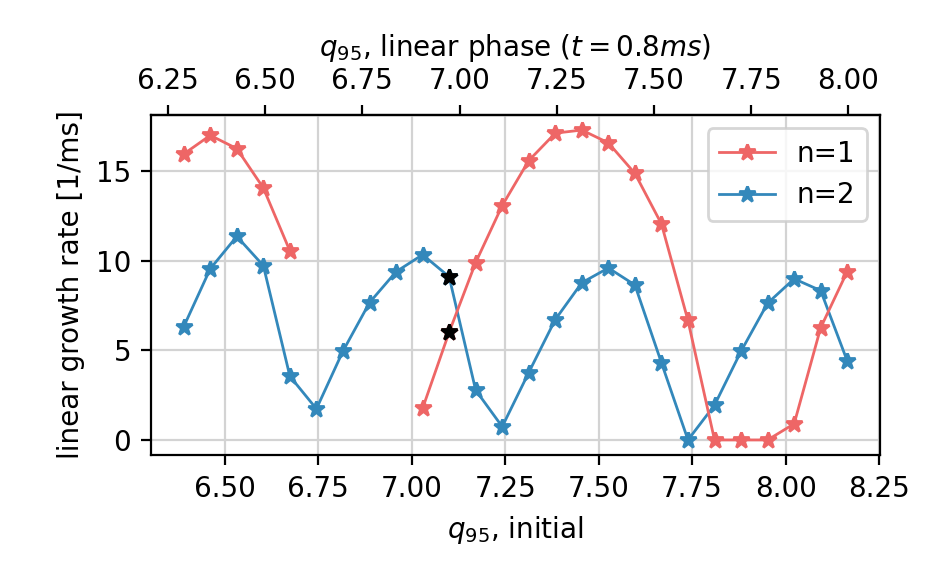}
	\caption[Linear growth rates as a function of $q95$]{Linear growth rates of the $n=1$ and $n=2$ modes as a function of initial $q_{95}$ (lower x-axis) and $q_{95}$ in the linear phase (upper x-axis). Note that between ${q_{95,\mathrm{init}} = 6.7}$ to $7.0$, numerical issues occurred in the simulations for $n=1$ such that this parameter range is excluded from the plot. The black stars represents the baseline case that has been analyzed in the previous sections.}
	\label{fig:qwin}
\end{figure}

Over the entire edge safety factor scan, only the $n=1$ and $n=2$ modes are linearly unstable, while perturbations with higher mode numbers are only non-linearly driven. In Figure~\ref{fig:qwin}, the linear growth rates are plotted as a function of the edge safety factor ($q_{95, \mathrm{init}}$) for the scanned window\footnote{The value of $q_{95}$ evolves during the simulation, therefore both the initial value of $q_{95}$ (in the lower x-axis), as well as $q_{95}$ in the linear phase at ${t = 0.8\si{\milli\second}}$ (in the upper x-axis) are indicated.}.

It is visible that the linear growth rates $\gamma$ as a function of $q_{95,\mathrm{init}}$ are two roughly sinusoidal functions with a period of $1$ for the $n=1$ mode and of $1/2$ for the $n=2$ mode, as is expected for external kink modes~\cite{Wesson_1978}. The $\gamma_{n=1}$ growth rates features three periods which correspond to a $m/n = 7/1$, $8/1$ and $9/1$ dominant KPM for $q_{95} \lesssim 6.9$, $6.9 \lesssim q_{95} \lesssim 7.9$ and $7.9 \lesssim q_{95} $ respectively. The linear growth rates of the $n=1$ and $2$ modes are slightly out of phase. Unfortunately, for the cases between ${q_{95, \mathrm{init}} = 6.7}$ and $7.0$, numerical issues \revised{originating at the magnetic axis} occurred in the simulations in the $n=1$ harmonic, hence the $n=1$ growth rates are missing from the plot (consequently, these simulations could not reach the non-linear phase either). 

Using the results from the linear scan alone, it is possible to identify parameter ranges of $q_{\mathrm{init}}$ in which no MHD activity occurs, however it is not possible to determine which cases will non-linearly develop KPMs forming a QH-mode state. Rather, it is necessary to continue the simulations into the non-linear phase. The following subsection describes such non-linear dynamics and the interactions of non-axisymmetric modes with the axisymmetric background. This provides insights into how the linear picture translates into the formation of QH-mode or bursting (ELM-like) behaviour.

\subsection{Non-linear evolution as a function of the edge safety factor} \label{sec:VariationQNonLin}

\begin{figure}[ht!]
	\centering
	\includegraphics[width=0.95\linewidth]{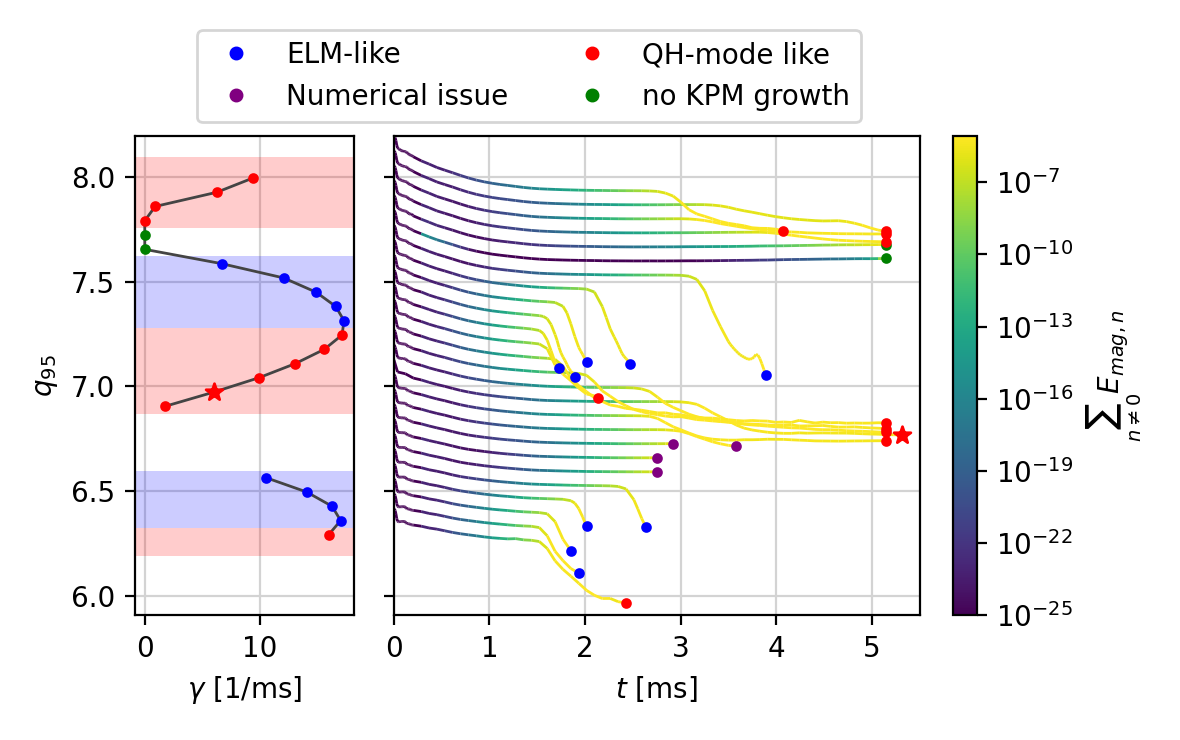}
	\caption[Evolution of $q95$]{Temporal evolution of $q_{95}$ for various values of ${q_{95, \mathrm{init}}}$ (right) and $\gamma_{n=1}$ as a function of $q_{95}$ in the linear phase (at \SI{0.8}{\milli\second}) (left). The color of the time traces indicates the total perturbed magnetic energy. The characteristic of each case is indicated by the color of the markers in the growth rate diagram on the left and at the end of each trace (or the cut-off at \SI{5.2}{\milli \second}) on the right. The cases marked red develop into QH-mode like state, blue indicates a strong ELM-like burst, green indicates stable solutions, and purple represents cases that were stopped due to numerical issues. To match a case in the left and right-hand side of the plot, the edge safety factor at \SI{0.8}{\milli\second} has to be considered. In the $n=1$ growth rate diagram, the approximate ELM and QH-mode windows are indicated by the background color. The star symbol represents the baseline case that has been analyzed in the previous sections.}
	\label{fig:qconv}
\end{figure}

The non-linear evolution for the different cases is displayed in Figure~\ref{fig:qconv}, which shows (in the right part of the figure) how the edge safety factor changes in time for all the different simulations performed in this scan, and links these different evolutions to the linear growth rate of the $n=1$ mode at \SI{0.8}{\milli\second} (shown in the left part of the figure). Initially, in the linear phase where the mode amplitudes are still small (lasting at least $1 \si{\milli\second}$ in all cases), ${q_{95}}$ evolves in a similar way for all cases due to slow profile changes resulting from source and diffusion terms. The subsequent evolution, however, is determined by non-linear physics and depends on the mode activity present in each simulation.

The non-linear evolution of cases with different ${q_{95, \mathrm{init}}}$ features three different types of solutions, namely QH-mode like cases, ELM-like bursts and MHD stable cases:

\begin{itemize}
\item QH-mode solutions are obtained for ${q_{95, \mathrm{init}} = 6.39}$, ${\{ 7.03 \dots 7.38\},}$ and ${\{7.95 \dots 8.16\}}$, where the modes saturate non-linearly and enter a quasi-stationary state. Particularly interesting is the fact that the QH-mode solutions non-linearly deviate from their initial value of edge safety factor due to the mode activity and converge to discrete values of $q_{95}$, which lie close to the minimal locations of the linear $n=1$ growth rate. This can be clearly observed in Figure~\ref{fig:qconv} where the cases converge to the discrete values of ${q_{95}\approx 5.9,}$ $6.8$, and $7.6$. This seems to be qualitatively consistent with observations from Ref.~\cite{Suttrop_2004}, which shows distinct QH-mode windows at $q_{95}=3.5\dots3.7$ and $4.4$ respectively.
\item Stable solutions are obtained for ${q_{95, \mathrm{init}} = \{7.81 \dots 7.88\}}$, where non-axisymmetric modes do not grow to cause visible changes to the axisymmetric background. This solution occurs when ${q_{95}}$ after $0.8~\mathrm{ms}$ is below the range of $q_{95}$ values with QH-mode\footnote{In the simulations at $q_{95}=6.8$ where ${q_{95}(\mathrm{at}~t=0.8~\mathrm{ms})}$ is also slightly smaller than the $q_{95}$ values with QH-mode, numerical issues appear.}. These stable cases would eventually lead to ELM-like crashes due to the continuous pedestal build-up.
\item Strong bursts are obtained for ${q_{95, \mathrm{init}} = \{ 6.46 \dots 6.67\}}$ and ${\{ 7.46 \dots 7.74\}}$, where the modes do not saturate but continue to grow and reach significantly higher magnetic energies than the QH-mode like cases and cause higher particle and heat losses. They take place when ${q_{95}(\mathrm{at}~t=0.8~\mathrm{ms})}$ is \revised{larger/}smaller than the discrete values of ${q_{95}}$ where the QH-mode solution is found, but not so \revised{large/}small that it approaches the next discrete value with QH-mode solution. For these cases, the burst event seems to bring the edge safety factor as well towards one of the discrete values where QH-mode is present. However, such violent cases are numerically challenging and were not continued long enough to assess whether they would also converge towards the discrete $q_{95}$ values and eventually form a QH-mode state after the initial MHD bursts.
\end{itemize}

The ranges of ${q_{95, \mathrm{init}}}$ for which the different solutions occur are observed to be such that QH-mode like cases occur where $\gamma_{n=1}$ reduces with decreasing ${q_{95,\mathrm{init}}}$ while ELM-burst like cases occur when $\gamma_{n=1}$ increases with decreasing ${q_{95,\mathrm{init}}}$. In the former case the drop in ${q_{95}}$ decelerates the mode growth, while for the latter the ${q_{95}}$ reduction accelerates the mode growth.

For the QH-mode cases, the influence of the mode on the evolution of the safety factor is hence found to be such that the plasma evolves towards a state with lower linear growth rate until it eventually reaches a state where kink modes become stationary. This proposes a possible saturation mechanism for the QH-mode.

The non-linear drop of $q_{95}$ is caused by an inward shift of the high pressure gradient region of the pedestal that moves the bootstrap current inwards. As a result, the total plasma current contained inside the $\psi_\mathrm{n}=0.95$ surface increases, leading to a reduction of the safety factor for this surface. 
The safety factor profile is most dominantly affected in the pedestal region between $\psi_\mathrm{n} \approx 0.875$  and  $\psi_\mathrm{n} \approx 0.975$. When analysing the ELM cycle simulations of Ref.~\cite{Cathey_2020}, it is possible to observe similar dynamics.

Depending on the initial value of $q_{95}$, the saturation amplitude varies among the cases which develop QH-mode. Generally, a higher linear growth rate translates into a higher saturation amplitude, which enhances the particle and heat transport. Note, that the non-linear evolution of $q_{95}$ towards these QH-mode windows seen in our simulations might not appear in the exact same way in the experiment, since the plasma current is usually feedback controlled there. Simulations with feedback control on the plasma current can be considered for future work. 

\begin{figure}[!ht]
	\centering
	\includegraphics[width=0.85\linewidth]{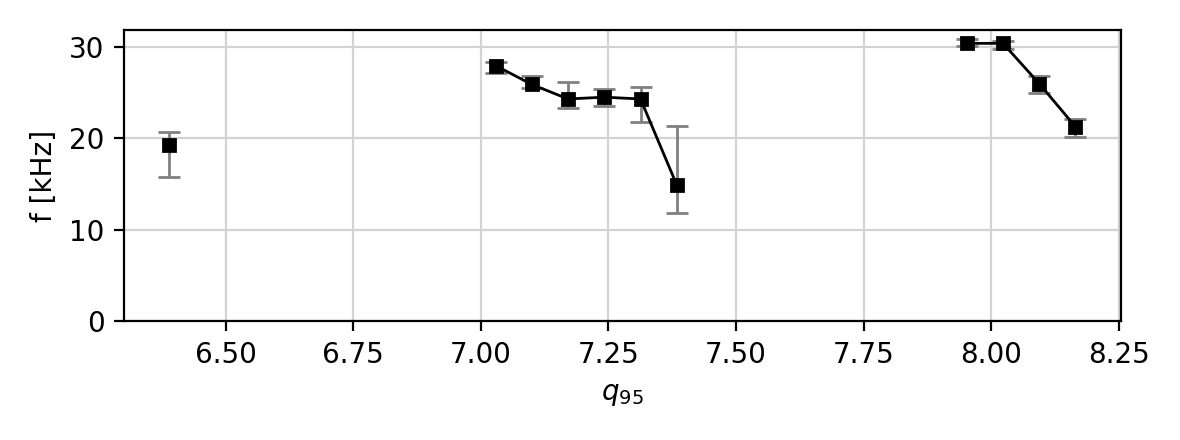}
	\caption[EHO rotation frequency as a function of $q_{95}$]{EHO rotation frequency as a function of initial $q_{95}$ for the cases identified as QH-mode. The error bars indicate the full-width half-maximum of the EHO peak in the spectra.}
	\label{fig:fvsq95}
\end{figure}

For cases which enter a QH-mode like state, the rotation frequency of the EHO was seen to depend on $q_{95}$ as displayed in Figure \ref{fig:fvsq95}.  Consistent with Ref.~\cite{Suttrop_2004}, it was observed that, within a QH-mode window, the EHO frequency decreases as $q_{95}$ increases. From one window to another, the frequency seems to show a weak increasing trend with $q_{95}$.

\section{Transition from QH-mode into an ELM regime}
\label{sec:QHtoELM}

To characterize the QH-mode regime, not only the access conditions under the variation of various parameters need to be studied as discussed in section~\ref{sec:lr8} and existing work in literature, but also the termination of QH-mode upon changing plasma parameters is of high relevance. In particular, it has been observed experimentally that the QH-mode regime is typically accessed at low densities and collisionalities. In the considered experimental scenario, the control of the density is challenging as discussed in section~\ref{sec:setup} and QH-mode terminates after a steady increase of density has occurred. It is, consequently, of significant interest to investigate the behaviour of QH-mode in the simulations when such an experimental density increase is reproduced.

Although it is technically possible to simulate the full experimental QH-mode phase lasting for 150 ms, the computational costs would be unreasonably high and the tearing mode occurring already much earlier in this case would complicate analysis. Thus, to reach the same pedestal top density values as found at the termination of QH-mode in the experiment more quickly, the particle source is increased in the following, to speed up the build-up of the density pedestal.

\begin{figure}
	\centering
	\includegraphics[width=1.0\linewidth]{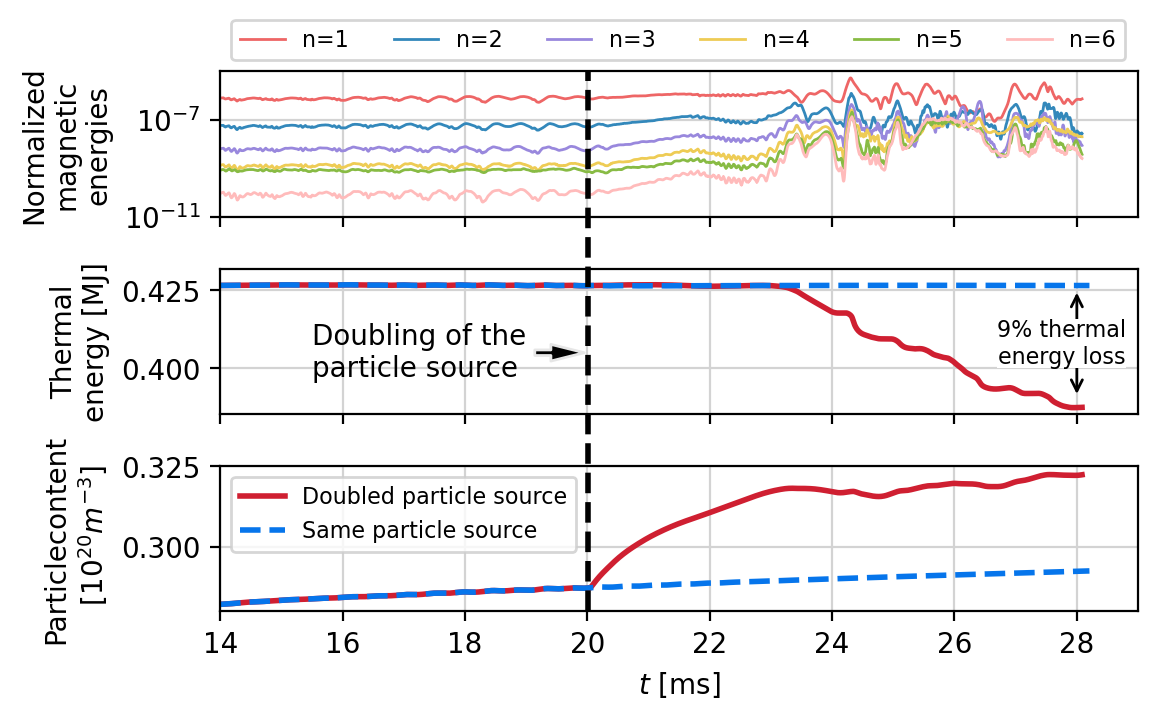}
	\caption[Dynamics after increasing the particle source]{Dynamics after increasing the particle source by a factor two at \SI{20}{\milli\second} compared to the nominal case. The magnetic energies, thermal energy and particle content are displayed compared to the evolution of the nominal case. The additional particle source immediately increases the particle content until an ELM-like crash occurs at about \SI{23}{\milli\second} and causes a significant loss of particles and heat. Over the course of the crash, around \SI{9}{\percent} of the thermal energy is lost.}
	\label{fig:macro_tr2}
\end{figure}

At first, we double the edge particle source compared to the previously shown cases\footnote{Resulting in an \SI{80}{\percent} increase of the particle source inside the separatrix, since the small core particle source is left unchanged} at \SI{20}{\milli \second}, that is about \SI{13}{\milli\second} after the formation of the quasi-stationary QH-mode in the simulation. The evolution of the magnetic energies, thermal energy and particle content of this case are displayed in Figure~\ref{fig:macro_tr2} and compared to the nominal case. It can be seen that before the change of the source, the plasma is in a quiescent steady state where the magnetic energy fluctuates only slightly and the particle content increases slowly, while the thermal energy stays roughly constant. When the particle source is increased, the magnetic energies, in particular for higher toroidal mode numbers, grow slowly at first until ${\approx23~\mathrm{ms}}$, when an ELM-like crash appears and removes about \SI{9}{\percent} of the thermal energy and brings the growth of the particle content to halt.

\begin{figure}[!ht]
	\centering
	\includegraphics[width=1.0\linewidth]{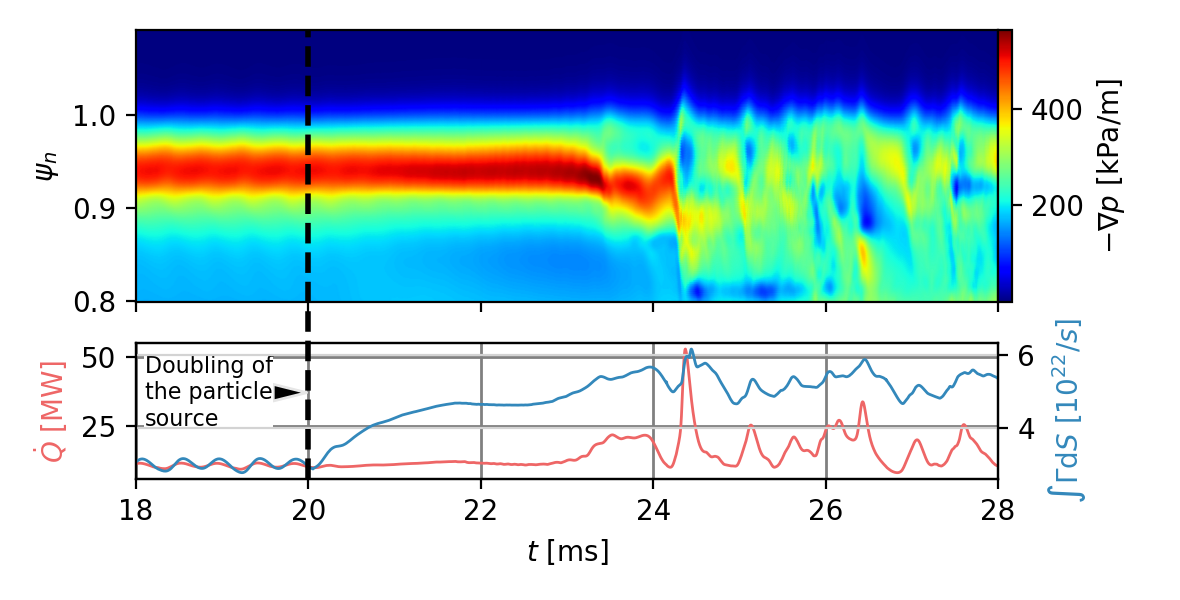}
	\caption[Collapse of the pedestal]{Collapse of the pedestal over the course of the ELM crash. The upper panel shows the evolution of the outboard midplane pressure gradient profile in the pedestal over time, the lower panel shows the boundary surface-integrated heat flux $\dot{Q}$, as well as the boundary surface-integrated particle flux ${\int \Gamma dS}$. The pedestal pressure gradient is greatly reduced during the ELM crash, while the heat flux out of the plasma is significantly increased. The particle flux already rises when the source is increased, but goes up further during the ELM crash.}
	\label{fig:ELM_p_pwr}
\end{figure}

Over the course of the crash most of the pressure pedestal collapses as depicted in Figure~\ref{fig:ELM_p_pwr}, which also displays the power flux from the plasma to the wall. Starting from the change in the density source, the pressure gradient increases slightly up to the point right before the onset of the crash at $23~\mathrm{ms}$. Thereafter, in the first phase of the crash until $24.5~\mathrm{ms}$, the lost power increases while the pressure gradient slightly reduces and shifts inwards. Following after this first smaller crash, at $24.5~\mathrm{ms}$ most of the pedestal pressure gradient is flattened within less than $0.2~\mathrm{ms}$, accompanied by a peak incident power of more than $50~\mathrm{MW}$.

\begin{figure}[!ht]
	\centering
	\includegraphics[width=0.95\linewidth]{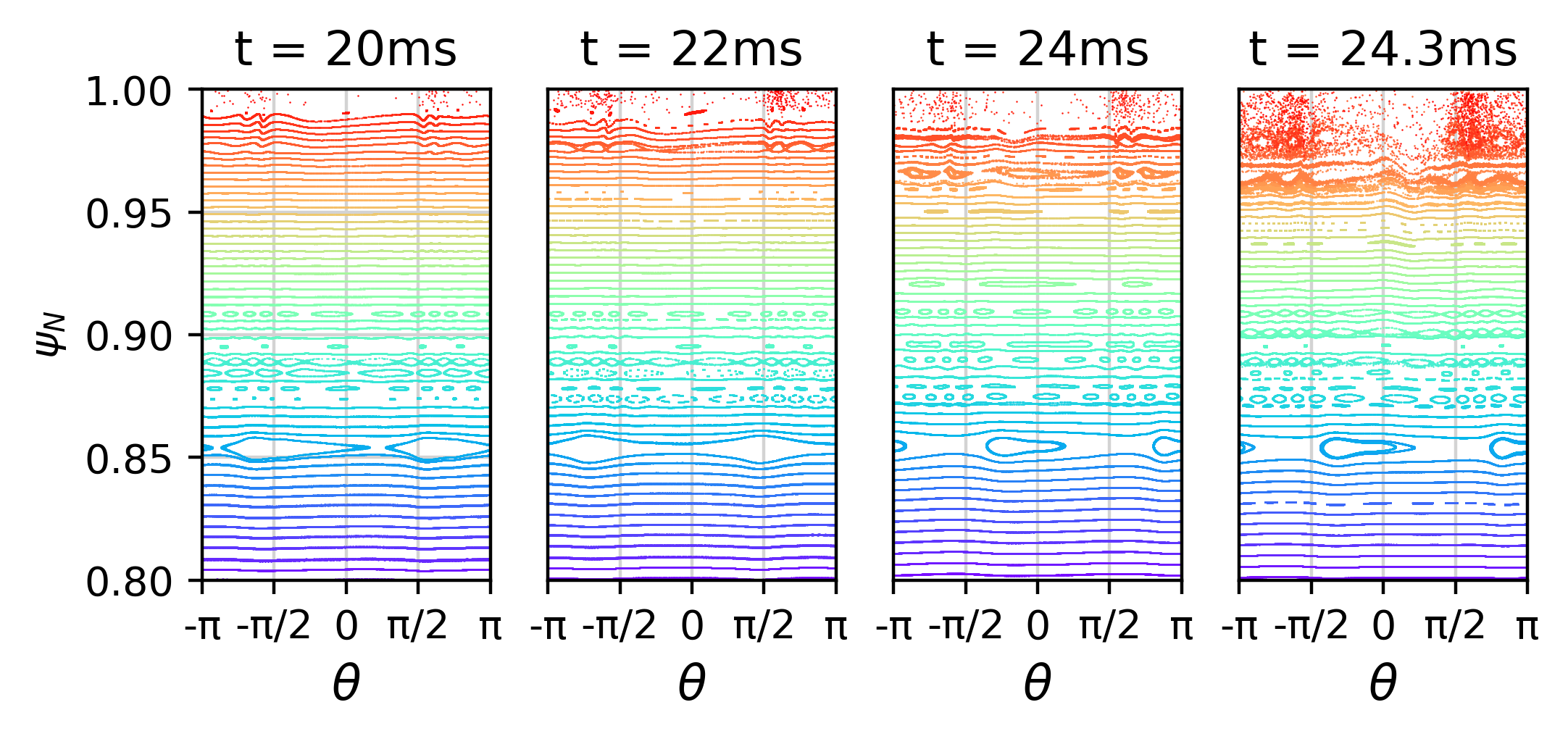}
	\caption[Poincaré plots during the crash]{Poincaré plots before the change of the source (a), before the onset of the ELM-like crash (b), during the start of the crash (c), and at the time of its peak power load (d).}
	\label{fig:tr2_Poincare}
\end{figure}

Poincaré plots at ${t=20.0}$, 22.0, 24.0, and ${24.3~\mathrm{ms}}$ are shown in Figure~\ref{fig:tr2_Poincare}. The first and second panel show that the change of the source does not have an immediate effect on the ergodisation of the field close to the separatrix. On the other hand, before the maximum power loss (in the early phase of the crash at ${t=24.0~\mathrm{ms}}$, shown in the third panel) the field becomes significantly more ergodised and closed flux surface only exist inside $\psi_\mathrm{n} \approx 0.975$. The field gets even more distorted during the highest ELM-induced power losses (fourth panel), where the last intact flux surface is at ${\psi_\mathrm{n} \approx 0.955}$, coherent with the large additional power flux to the wall (Figure~\ref{fig:ELM_p_pwr}). Further inside, several magnetic islands are present featured, most notably a ${m/n=2/1}$ tearing mode.

\begin{figure}[!ht]
	\centering
	\includegraphics[width=0.8\linewidth]{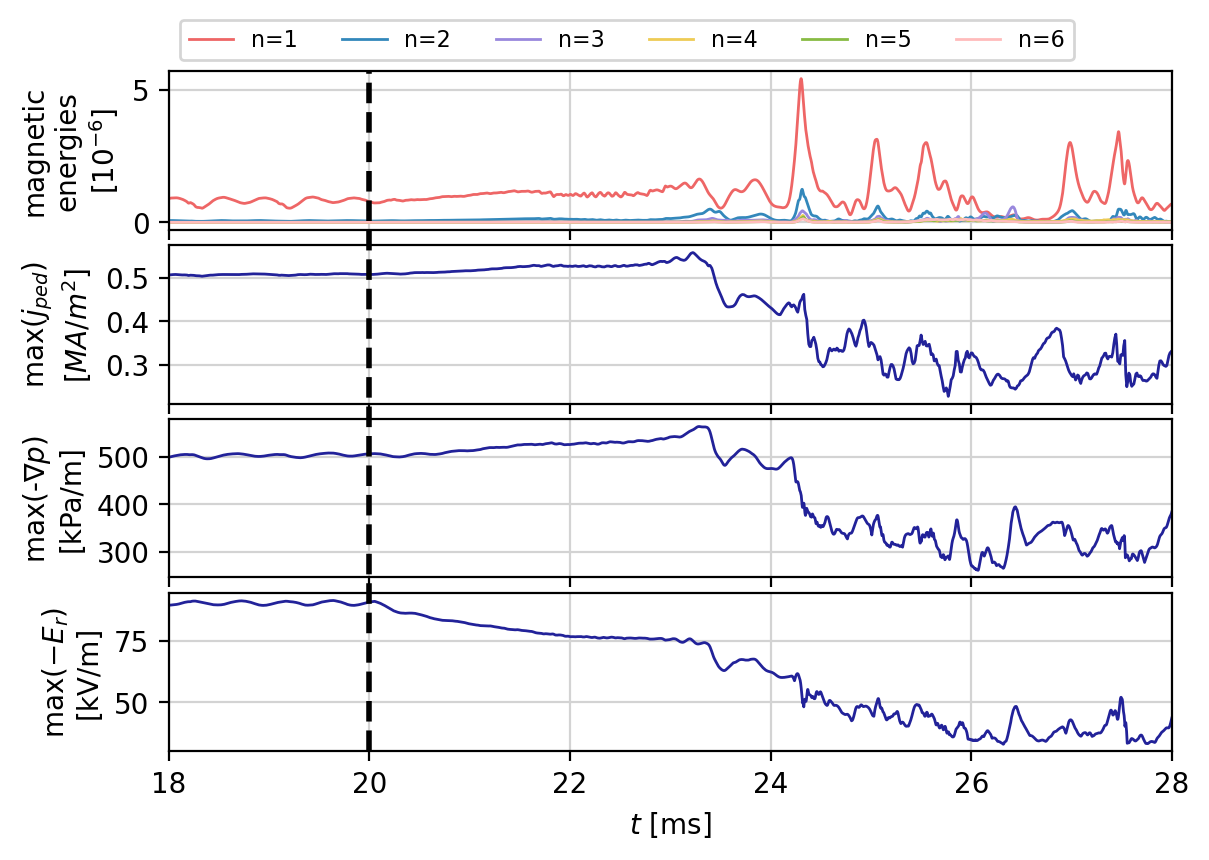}
	\caption[Stability contributions during the crash]{Evolution of the magnetic energy before and during the ELM crash. For the destabilizing current density $j$, the maximum of the flux-surface averaged pedestal profile is given. The destabilizing pressure gradient $\nabla p$ and the stabilizing radial electric field $E_r$ contributions are indicated by their extremal values in the toroidally averaged outer midplane.}
	\label{fig:Crash_Stability}
\end{figure}

In Figure~\ref{fig:Crash_Stability}, the magnetic energy evolution is shown again (in a non-logarithmic plot, see Figure~\ref{fig:macro_tr2} for the logarithmic representation), together with the most important quantities that play a role in the stability of the pedestal, namely the maximum current density, the maximum pressure gradient and the minimum of the radial electric field well. In this simplified analysis, we neglect that stability will of course also depend on the relative radial location, the pedestal width, etc. It can be seen that the pressure gradient, as well as the pedestal current, increase moderately while the electric field well depth is being reduced right away when the particle source is turned on, as expected due to $E_r\propto 1/n_e$~\cite{Viezzer_2014}.

Critically, this effect reduces the stabilising influence of flows onto ballooning modes and, at the same time, increases the destabilising terms for both ballooning and peeling modes. As visible in the logarithmic representation of the magnetic energies in Figure~\ref{fig:macro_tr2}, at first, a significant growth of the $n \geq 2$ mode and higher harmonics sets in at \SI{23}{\milli \second}. Thereby, the contributions of the higher $n$ modes become more important, e.g., the $n=6$ mode increases in amplitude by more than three orders of magnitude while the dominant $n=1$ mode even decreases slightly in the early phase of the crash and only increases by a factor $5$ in the most violent phase. This indicates the transition from a saturated KPM to a transient burst of peeling-ballooning activity, leading to the abrupt collapse of the pedestal and a large peak in the boundary heat flux. 

The increase in density and part of the crash were repeated with higher toroidal resolution in a separate simulation. The onset point of the crash is not affected and the qualitative dynamics of the ELM crash are captured by the simulation, however the detailed dynamics of the crash are not fully resolved. We expect the temporal scales of the crash to be faster with higher toroidal resolution as reported in Ref.~\cite{Cathey_2020}. For this reason, the details of the ELM crash should be only considered qualitatively, while the onset point of the ELM is a quantitatively robust result. 

\begin{figure}[!ht]
	\centering
	\includegraphics[width=0.7\linewidth]{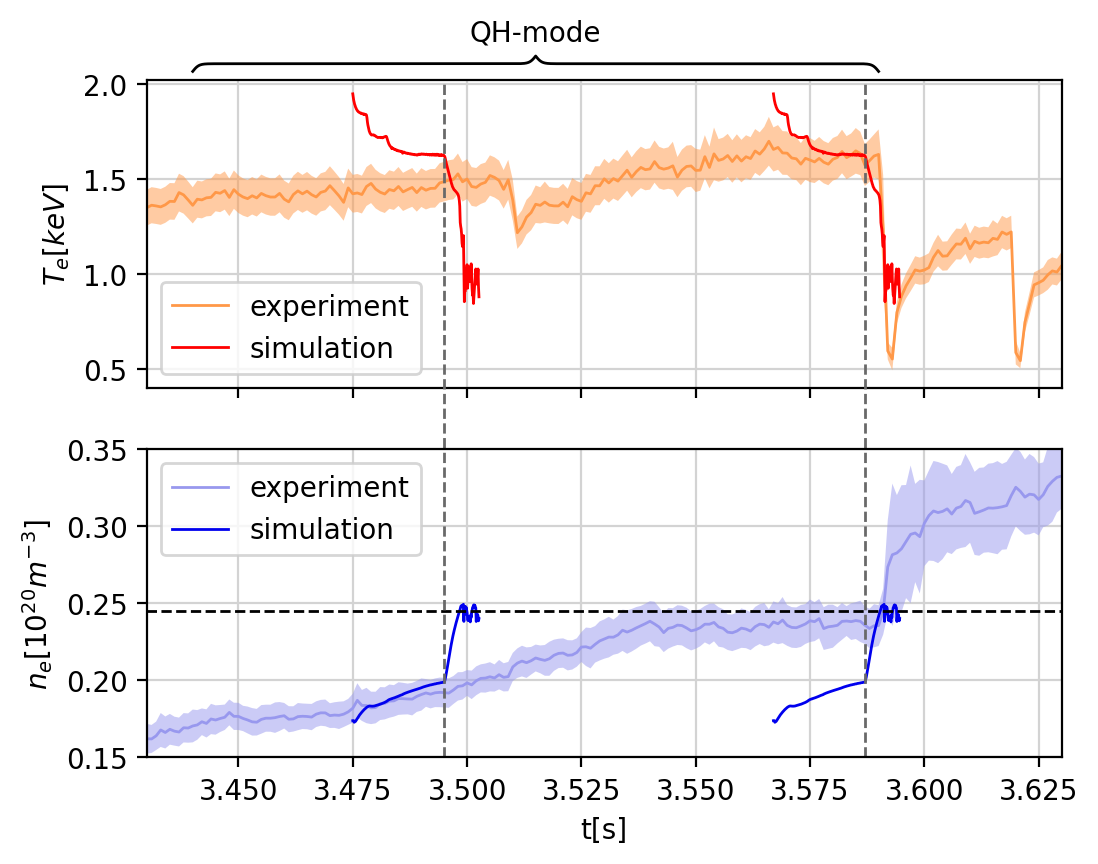}
	\caption[Pedestal top density and temperature]{Evolution of the density and temperature at the pedestal top ($\psi_\mathrm{n} = 0.95$) in simulation and experiment. The simulation trace is shown twice -- once aligned to its starting point (left) and once aligned to the ELM crash (right). The vertical dashed lines indicate the change of the particle source while the horizontal dashed line indicates the density value at which QH-mode is lost in the simulation, matches the experimental observation within the error bars.}
	\label{fig:pedestaltopexpsim}
\end{figure}

Indeed, it is observed that the conditions that lead to the loss of QH-phase in the simulation compares remarkably well with the experimental excursion away from the QH-phase at ${t_\mathrm{exp.}=3.59~\mathrm{sec.}}$ (see fig.~\ref{fig:spectrogram39279}). This is shown in Figure~\ref{fig:pedestaltopexpsim}, which displays the evolution of the density and temperature at the pedestal top ($\psi_\mathrm{n} = 0.95$). The experimental data obtained from IDA~\cite{Fischer_2010} is shown for part of the QH-mode phase and the first two ELMs after its termination. For better comparability, the trace of the simulation is shown twice. On the left side it is aligned such that the start of the simulation matches the time of the equilibrium reconstruction, in this trace the evolution of density and temperature before the change of the source approximately matches the experimental one. On the right, the simulation is shifted in time such that the time of the crash coincides with the time of the first ELM in the experiment. The black dashed line indicates the density at which the ELM crash is triggered in the simulation, which matches the experimentally observed density within the error bars.

At \SI{3.59}{\second}, slightly delayed with respect to the drop in temperature, the experimental density can be seen to rise substantially, which is a feature not captured by the simulation. Likely this sudden increase of the pedestal density is due to recycling and ionisation of neutrals in the SOL/divertor region resulting from the large power flux onto the plasma facing components caused by the ELM crash. As the model used here represents the SOL in a simplified way, such effects are not incorporated and, therefore, such response cannot be expected. 

A second case, which has the edge particle source increased by only \SI{88}{\percent} instead of \SI{100}{\percent}, confirms that the critical density at which the QH-phase is lost is not determined by the change in particle source. This second case also evolves into an ELM crash, of which only the very beginning was simulated, occurring at a \SI{4}{\percent} lower pedestal top density than the case with a doubled source which is also well within the error bars of the experimental measurement. 

These simulations show for the first time that it is possible to reproduce the transition from QH-mode into an \revised{ELM crash} as a result of a density increase. The effect of the density on the electric field is proposed to be the critical contribution to the destabilisation of the pedestal, which leads to the onset of an ELM crash. Despite the shortcomings in terms of the simulation of the ELM crash itself, the most important feature, namely the critical pedestal top density, corresponds to the experimental density at which the QH-phase is lost, and gives confidence that the access and loss conditions of QH-mode can be predicted with the MHD model used here.

\section{Limit Cycle oscillations}
\label{sec:LCO}

In section \ref{sec:QHtoELM}, the QH-mode regime was lost when a critical density value was reached. At the critical value, the pedestal became sufficiently unstable such that an ELM crash was triggered. In said simulations, the density evolution had been accelerated beyond \revised{that of section~\ref{sec:lr8}} in order to save computational costs by doubling the particle source. In the present section, a phenomenon is described that occurs when the particle source is increased more moderately, such that the pedestal successively becomes more unstable, but does not trigger an ELM crash right away. Instead, an oscillation in the pedestal with a different nature than the already present EHO, emerges. 

\begin{figure}
	\centering
	\includegraphics[width=0.9\linewidth]{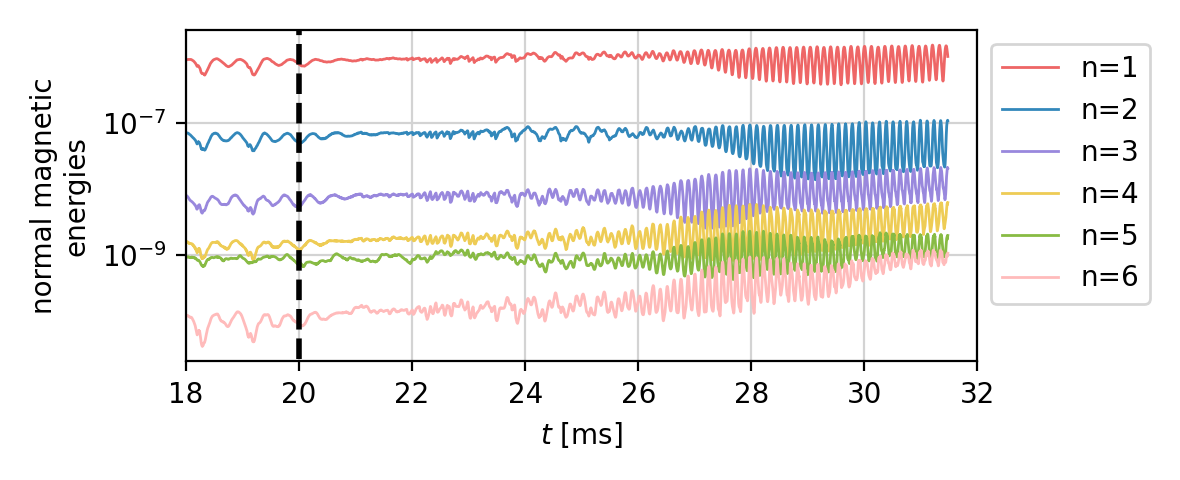}
	\caption[Magnetic energy oscillations during the onset of the limit cycle]{Magnetic energy oscillations during the onset of the limit cycle. The particle source is increased at \SI{20}{\milli\second} by \SI{22}{\percent}, marked with a dashed line. After about \SI{5}{\milli\second}, the oscillations can be seen to emerge and increase in size, reaching their full amplitude at \SI{28}{\milli \second}.}
	\label{fig:LC_magE}
\end{figure}

\begin{figure}[!hp]
	\begin{subfigure}{0.49\linewidth}
		\centering
		\includegraphics[width=1.0\linewidth]{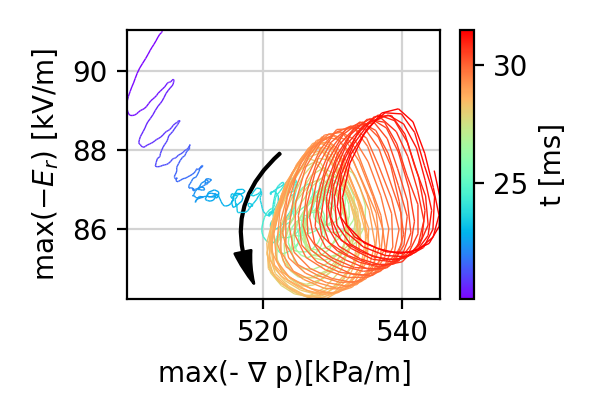}
		\caption[]{}
		\label{fig:LCO_PErt}
	\end{subfigure}
	\hfill
	\begin{subfigure}{0.49\linewidth}
		\centering
		\includegraphics[width=1.0\linewidth]{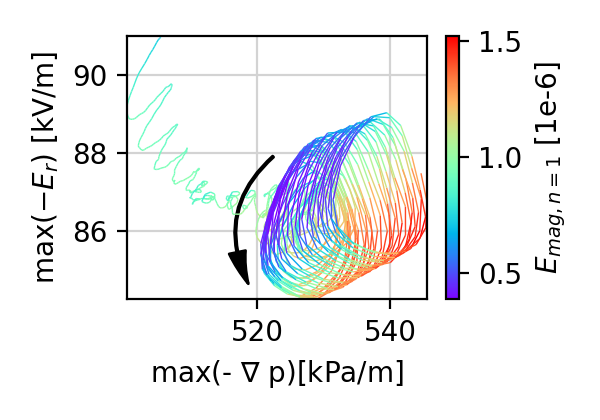}
		\caption[]{}
		\label{fig:LCO_pErE}
	\end{subfigure}
	
	\begin{subfigure}{0.49\linewidth}
		\centering
		\includegraphics[width=1.0\linewidth]{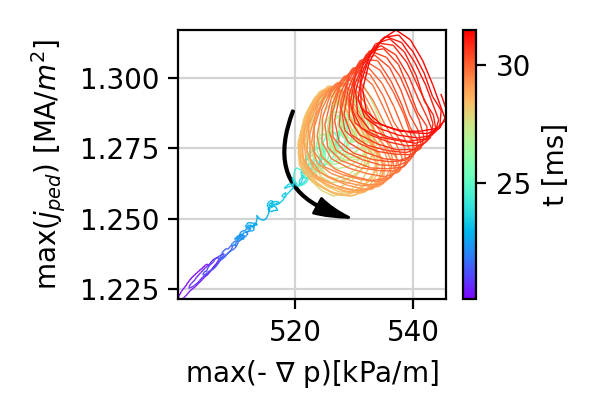}
		\caption[]{}
		\label{fig:LCO_pjt}
	\end{subfigure}
	\hfill
	\begin{subfigure}{0.49\linewidth}
		\centering
		\includegraphics[width=1.0\linewidth]{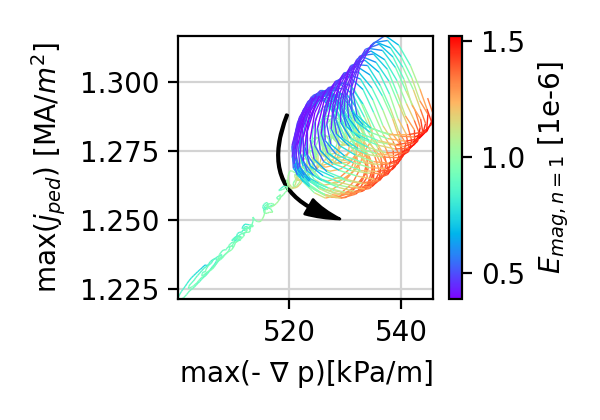}
		\caption[]{}
		\label{fig:LCO_pjE}
	\end{subfigure}
	
    \begin{subfigure}{0.49\linewidth}
		\centering
		\includegraphics[width=1.0\linewidth]{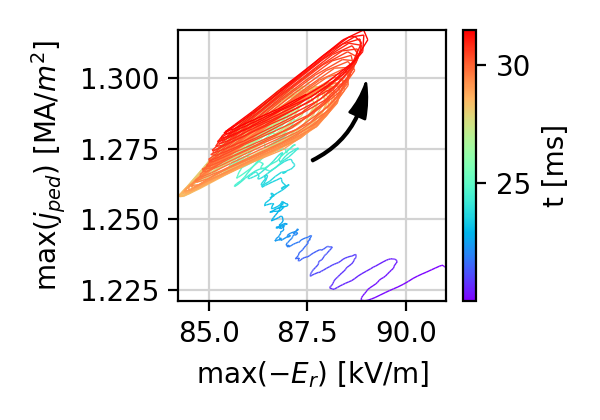}
		\caption[]{}
		\label{fig:LCO_Erjt}
	\end{subfigure}
	\hfill
	\begin{subfigure}{0.49\linewidth}
		\centering
		\includegraphics[width=1.0\linewidth]{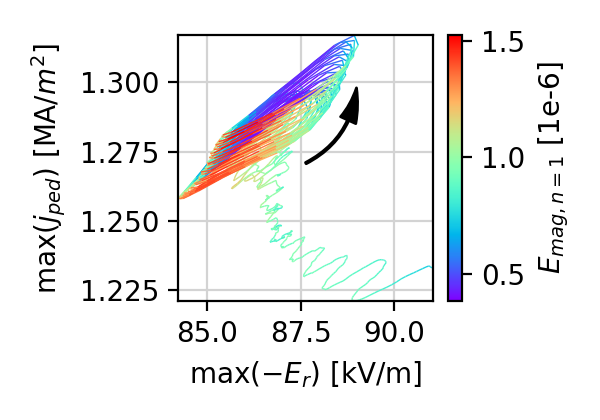}
		\caption[]{}
		\label{fig:LCO_ErjE}
	\end{subfigure}
	\caption[Limit cycle oscillations in phase space]{Limit cycle oscillations in phase space, colored by time (\ref{fig:LCO_PErt}, \ref{fig:LCO_pjt} and \ref{fig:LCO_Erjt}) and colored by the normalized magnetic energy of the $n=1$ mode (\ref{fig:LCO_pErE}, \ref{fig:LCO_pjE} and \ref{fig:LCO_ErjE}). From \SI{20}{\milli \second} to \SI{25}{\milli \second}, the pedestal drifts to a higher pedestal with a lower radial electric field, that is towards a more unstable state. From \SI{25}{\milli \second}, the cycle starts to establish and reaches its full size at \SI{28}{\milli \second}. Thereafter the limit cycle maintains its size, but drifts slowly over time due to the sourcing of the pedestal. It can be seen that the oscillations of the magnetic energy is in phase with the limit cycle.}
	\label{fig:LCO}
\end{figure}

Similar to the case presented in section~\ref{sec:QHtoELM}, the edge particle source was increased at \SI{20}{\milli \second}, however here it was only increased by \SI{22}{\percent}. After about \SI{6}{\milli\second}, a global oscillation emerges, during which the entire pedestal flow periodically accelerates and decelerates. At the same time, the amplitude of the saturated KPM starts to \revised{become modulated}, as can be seen in the magnetic energies in figure~\ref{fig:LC_magE}. At a frequency of \SI{6}{\kilo \hertz}, this oscillation is about a factor of four slower than the EHO, which remains present also when the additional pedestal oscillation forms. The oscillations of the background kinetic energy is phase inverted compared to the $n\neq0$ kinetic energy contributions, while the latter is in-phase with the $n\neq0$ magnetic energy oscillations. The effect of the oscillations is also observed in the pedestal radial electric field, current, pressure, density and temperature. All said quantities fluctuate on the order of \SI{5}{\percent} around a slowly evolving value which can be clearly seen in Figure~\ref{fig:LCO}. While the influence of the EHO onto the axisymmetric background results from the rotation of a helical perturbation, this additional oscillation rather constitutes a change of the toroidally averaged quantities (i.e., the $n=0$ components).

As an indicator for the stability of the pedestal, the maximum values of radial electric field, current and pressure gradient profiles can be investigated. The quantities can be plotted in phase space to indicate the stability of the pedestal, which is displayed in Figures~\ref{fig:LCO_PErt}, \ref{fig:LCO_pjt} and~\ref{fig:LCO_Erjt}. It is visible that already initially after the change of the source, small oscillations were present in the pedestal. After about \SI{5}{\milli \second} though (i.e., $t\geq25\si{\milli\second}$), a far larger and quasi-periodic pedestal oscillation emerges that almost closes on itself after each period in phase space. On top of the periodic oscillation, a slow drift towards higher pressure gradients, current density and electric fields is visible that is caused by the slow build up of the pedestal. By coloring the trajectory of this limit cycle oscillation according to the $n=1$ magnetic energy in Figures~\ref{fig:LCO_pErE}, \ref{fig:LCO_pjE} and~\ref{fig:LCO_ErjE}, it can be seen that the mode grows and shrinks in phase with the oscillation. As the mode grows and shrinks in amplitude, it is thought that the stabilizing and destabilizing factors change in relative magnitude over the course of an oscillation cycle. 

During the onset phase of the oscillation, the oscillation amplitudes of current, pressure gradient and radial electric field grow and seem to converge towards an almost closed cycle in phase space with a particular amplitude. Thereafter, the oscillation amplitudes stop growing and the oscillation remains on almost the same path in phase space, only overlaid with a small drift due to the build up of the pedestal. Due to this observation of convergence to a certain oscillation amplitude, the phenomenon was dubbed to be a limit cycle oscillation. 

This kind of oscillation was seen in all cases where the particle source was increased after the QH-mode had established (in the nominal case, this would likely emerge as well, but only after a very long time). However, the larger the particle source, the faster the pedestal evolves. When the time scale of the density build-up gets comparable to the one of the limit cycle oscillation, no clear quasi-periodic oscillation is forming any more (like in the case discussed in the previous section with 100\% increase of the edge particle source). With \SI{44}{\percent} increase of the source, the drift becomes faster compared to the 22\% case, but the limit cycle can still establish properly. For cases with \SI{66}{\percent}, \SI{88}{\percent} or \SI{100}{\percent} edge particle source increase, however, only very few cycles occur, before the onset of the pedestal collapse. Over these few cycles, the oscillation amplitude is still growing and cannot reach the fully saturated amplitude.

The maximal amplitude of the established oscillation in magnetic energy is about \SI{10}{\percent} larger for the case with \SI{44}{\percent} particle source increase compared to the previously mentioned case with a 22\% increase, thus, the amplitude of the oscillation depends on the drive. When reducing the particle source again after the limit cycle has established, it can further be seen that the magnitude of the oscillations begins to shrink just after the change of the source, following the reduced drive.

We have shown here a limit cycle oscillation close to the onset point of an ELM crash, in which the mode amplitude periodically varies. Consequently, the transport caused by the mode is varying periodically over time as well, such that the equilibrium profiles of density, temperature, radial electric field, and current density vary in phase with the amplitude variation. The interaction between the profile changes, which change the stability properties of the mode, and the transport caused by the mode are leading to the observed quasi-periodic oscillations. 

Similar oscillations to the ones simulated here are also reported in the literature of QH-mode experiments in JT60-U~\cite{Kamiya2021}. During the QH-mode phase, the plasma is observed to be in a dynamical steady-state where the dominant peeling mode undergoes repeated growth and damping cycles. Together with the fundamental mode of the EHO, the mean temperature gradient was measured to oscillate in a limit cycle with an LCO frequency about a factor $60$ lower than the EHO frequency. A limit-cycle model for these oscillations in JT60-U was developed in Reference \cite{Itoh_2021}. 

With this model it was shown that a limit cycle solution exists near the stability boundary of peeling modes, in which the pressure gradient and peeling mode amplitude oscillate in time, quantitatively consistent with the JT60-U observations. The experiment and the limit-cycle model shares qualitative characteristics, in particular the oscillating global mode amplitude, with the LCO found in our study.

\section{Conclusions}\label{sec:conclusions}

In this article, a Quiescent H-mode in the all tungsten ASDEX Upgrade was simulated. The experimental scenario, on which these simulations are based, shows a QH-mode phase, obtained in an upper single null divertor configuration without pumping~\cite{Viezzer_NF_2023}, lasting for 150 ms, during which the density rises continuously leading eventually to the loss of the QH-mode phase and the occurrence of periodic ELM crashes. The simulations performed manage to sustain an EHO, and observe a transition to a type-I ELM upon increasing the pedestal density towards the value where the QH-mode phase is lost in the experiment.

The simulations were set up according to the experimental equilibrium reconstruction and the heat and particle sources, as well as the perpendicular diffusion coefficients were set up to \revised{maintain the experimental temperature pedestal constant and to evolve the density pedestal at a faster rate than the experiment. This deliberate choice follows from the approach of Ref.~\cite{Cathey_2020} in which repetitive type-I ELMs are simulated}. In the nominal simulation, the $n=1$ and $n=2$ modes are unstable, while the higher modes are stable, but eventually become non-linearly driven. The $n=2$ mode has a larger linear growth rate and saturates first, while the $n=1$ mode saturates somewhat later, becoming non-linearly dominant at a higher amplitude. After that, a quasi-stationary spectrum of modes develops, which remains stationary for the duration of the simulated time of \SI{20}{\milli\second} after the saturation. The coupled modes cause a rigidly rotating non-sinusoidal density perturbation reminiscent of the experimentally observed edge harmonic oscillation.

A detailed look at the dependency of the dynamics on the initial value of $q_{95}$, varied in a set of simulations by changing the toroidal magnetic field amplitude, reveals that the linear growth rates of $n=1$ and $2$ vary between ${\gamma_{n=1}\sim0 \dots 1.6\cdot10^4~\mathrm{s}^{-1}}$ and ${\gamma_{n=2}\sim0 \dots 1\cdot10^4~\mathrm{s}^{-1}}$, respectively, with a 1-periodic structure in $q_{95}$ of $\gamma_{n=1}$ and a 0.5-periodic structure of $\gamma_{n=2}$. Depending on the initial value of $q_{95}$, the immediate emergence of QH-mode is observed in certain windows, while other values lead to an ELM-like burst. All cases seem to non-linearly evolve in $q_{95}$ towards discrete values, which is qualitatively consistent with experimental observations of $q_{95}$ windows for QH-mode formation in AUG-C~\cite{Suttrop_2004}. The QH-mode solutions occur in the $q_{95}$ region where the linear growth rate decreases with $q_{95}$ whereas ELM-like solutions are obtained when the growth rate increases with $q_{95}$. This change of the edge safety factor, towards a marginally stable state, seems to be an important, not previously described, saturation mechanism that regulates the amplitude of the saturated kink peeling modes of the QH-mode. 

The MHD simulations modelling the density rise occurring in the experiment show a loss of the QH-mode regime with the occurrence of an ELM crash: medium-n toroidal modes grow in addition to the stationary kink-peeling modes spectrum, leading to an ELM-like bursting dynamics in which significant amounts of heat and particles are expelled from the plasma. For numerical efficiency, the simulated \revised{rate at which the pedestal} density rises was increased compared to the density increase in the experiment. The decrease of the flow stabilization with increasing density seems to play a key role for the transition from the stationary state into the crash. The onset of this crash is in agreement with the experimentally observed density threshold. Further simulations, in which the time scale of the density rise is closer to the experimental behaviour, show a limit cycle oscillation, during which the equilibrium profiles of density, temperature, current density, and radial electric field change periodically and in phase with the amplitude of the $n=1$ kink peeling mode. 

Generally, the simulations presented here are able to reproduce the experimental observations from ASDEX Upgrade. Several aspects are in quantitative agreement, others match at least qualitatively. This work, thus, contributes both to a detailed understanding of the mechanisms associated with QH-mode, and to the validation of the JOREK code against experiments as it is needed for building confidence in predictive simulations. 

Future work will need to concentrate on further improving the models used for such simulations and pursuing further detailed comparisons to the experiment. Of particular interest might be, that the simulations show a higher density perturbation on the HFS than on the LFS, that the simulations predict specific discrete values of $q_{95}$ in the established QH-mode, and that the simulations predict limit cycle oscillations close to the termination of the QH-mode. \revised{The role of rotational shear is known to be an important player in QH-mode~\cite{Burrell_2001,Burrell_2009PRL}. The present contribution simplified the picture by neglecting toroidal rotation source, thereby including rotational shear only through the diamagnetic drift (which in the simulations sets the $E_r$ well). Future work on the QH-mode in AUG will focus on investigating the role of rotational shear by including the experimentally-relevant source of toroidal rotation. In doing so, the EHO frequency obtained in the simulations should show a better match to what is observed experimentally.}

Some aspects of the simulations will need to be improved in future work, to capture some of the details even more accurately. This includes the use of a two-temperature model, a slight reduction of resistivity to become fully realistic or, even better, a resistivity profile that considers the neoclassical modifications to the Spitzer resistivity. Furthermore a detailed assessment of the role of viscosity, a free boundary treatment that removes the ideal wall assumption at the boundary of the computational domain, a more accurate model for scrape-off layer and divertor including kinetic neutrals could be considered. If the ELM crash of the QH-mode to ELM transition should be resolved in detail, also a higher toroidal resolution would be needed.

\section*{Acknowledgements}

This work has been carried out within the framework of the EUROfusion Consortium, funded by the European Union via the Euratom Research and Training Programme (Grant Agreement No 101052200 — EUROfusion). Views and opinions expressed are however those of the authors only and do not necessarily reflect those of the European Union or the European Commission. Neither the European Union nor the European Commission can be held responsible for them. Some of the simulations were carried out on the Marconi-Fusion supercomputer operated by CINECA in Italy and on the Raven and Cobra systems operated by MPCDF in Germany.
E.V. and D.J.C.Z. gratefully acknowledge funding from the European Research Council (ERC) under the European Union’s Horizon 2020 research and innovation programme (grant agreement No. 805162) and A.C. from an EUROfusion Researcher Grant.

\printbibliography

\end{document}